\documentclass[journal,twocolumn,final]{IEEEtran} 

\usepackage[utf8]{inputenc}
\usepackage[T1]{fontenc}
\usepackage[hidelinks]{hyperref}
\usepackage{url}
\usepackage{ifthen}
\usepackage{cite}
\usepackage[cmex10]{amsmath} 
\usepackage{amsfonts,amssymb}
\usepackage{graphicx}
\usepackage{color}
\usepackage{ifthen}
\usepackage{comment}

\usepackage{tikz,pgfplots}
\pgfplotsset{width=7cm,compat=1.3}
\usepackage[justification=centering]{caption}

\usepackage[arxiv]{optional}  

\allowdisplaybreaks[1]

\def\cA{\mathcal{A}}

\def\cT{\mathcal{T}}

\newtheorem{lemma}{Lemma}

\newtheorem{theorem}{Theorem}

\newcommand{\E}{\mathbb{E}}

\def\P{\mathbb{P}}
\DeclareMathOperator {\qd}{qd}
\DeclareMathOperator {\tr}{tr}

\usepackage{mathtools}

\newcommand{\cN}{\mathcal{N}}

\newcommand{\cH}{\mathcal{H}}

\newcommand{\hv}{\hat{v}}
\newcommand{\chv}{\hat{V}}
\newcommand{\hhv}{\tilde{v}}
\newcommand{\vc}[1]{{#1}}
\newcommand{\mat}[1]{\mathbf{#1}}

\def\P{\mathbb{P}}

\title{Structured Equilibria for Dynamic Games with Asymmetric Information and Dependent Types}

\author{Nasimeh Heydaribeni and Achilleas Anastasopoulos%
	\thanks{This work was supported in part by NSF Grant ECCS-1608361.}
	\thanks{This work has been partially presented in \cite{HeAn19}.}
	\thanks{The authors are with the Department of Electrical Engineering and Computer Science, University of Michigan, Ann Arbor, MI, 48105 USA {\tt\small {heydari,anastas}@umich.edu}}
}

\begin{document}

	\maketitle
	\thispagestyle{empty}
	\pagestyle{empty}
	

	\maketitle
	
	\begin{abstract}
		We consider a dynamic game with asymmetric information where each player observes privately a noisy version of a (hidden) state of the world $V$, resulting in dependent private observations. We study structured perfect Bayesian equilibria that use private beliefs in their strategies as sufficient statistics for summarizing their observation history. The main difficulty in finding the appropriate sufficient statistic  (state) for the structured strategies arises from the fact that players need to construct (private) beliefs on other players' private beliefs on $V$, which in turn would imply that an infinite hierarchy of beliefs on beliefs needs to be constructed, rendering the problem unsolvable.
		We show that this is not the case: each player's belief on other players' beliefs on $V$ can be characterized by her own belief on $V$ and some appropriately defined public belief. We then  specialize this setting to the case of a Linear Quadratic Gaussian (LQG) non-zero-sum game and we characterize linear structured PBE that can be found through a backward/forward algorithm akin to dynamic programming for the standard LQG control problem.
		Unlike the standard LQG problem, however, some of the required quantities for the Kalman filter are observation-dependent and thus cannot be evaluated off-line through a forward recursion.
	\end{abstract}
	

	\section{Introduction}

	Dynamic games with asymmetric information play an important role in decision and control problems, yet there is no general framework to study such games in a tractable manner. The appropriate solution concept for these games is some notion of equilibrium such as Bayesian Nash equilibrium, perfect Bayesian equilibrium (PBE), sequential equilibrium, etc.~\cite{OsRu94, FuTi91,watson2017general}.
	Due to the dynamic nature of such games, the players' histories expand with time and therefore the corresponding strategies have an expanding domain. To mitigate this problem, researchers have introduced equilibrium concepts that summarize the time expanding histories into sufficient statistics. For symmetric information games Markov perfect equilibria~\cite{maskin2001markov} have been introduced, in which the players' strategies depend only on payoff-relevant past events and not the whole history. For asymmetric information games or control problems, finding the appropriate sufficient statistic is a challenging task and various information structures and corresponding statistics have been considered in the literature~\cite{OuTaTe17, vasal2018systematic, mahajan2015sufficient,yuksel2009stochastic,tavafoghi2018unified}.
	
	A quantity commonly used as a sufficient statistic, is a belief over some unknown part of the system.  The main challenge in this context is the emergence of private beliefs in the sufficient statistics, i.e., the fact that different agents in the system may have different (private) observations about the same quantity. One way to avoid this problem is to consider models in which private beliefs either do not exist (symmetric information games, or asymmetric but independent observations~\cite{OuTaTe17,VaAn16a, vasal2018systematic}) or, if they exist, they are not taken into account in agents' strategies (see for example the concept of ``public perfect equilibrium''~\cite{abreu1990toward}).  In order to intuitively explain the conceptual difficulty arising from having private beliefs in the sufficient statistics, consider the following thought process. If a player acts according to her private belief $\xi^i_t$ of a hidden variable and she expects other players to behave in the same way, she needs to form a belief over other players' beliefs to interpret and predict their actions and she has to take that belief into account when acting. In other words, she has to form a belief over (at least) $\xi^j_t$ for all other users $j\neq i$. This is a belief on beliefs which is also a private information of user $i$ and it has to be taken into account in her strategies. Due to symmetry of the information structure, all other players should do the same. But now, it is clear that user $i$ needs to form beliefs over beliefs over beliefs of other players.
	This chain continues as long as this hierarchy of beliefs are private. It stops whenever the beliefs in one step are public or public functions of previous step beliefs.
	
	In this paper, we study a dynamic game with asymmetric information. We consider a model with an unknown state of the world $V$, where each player $i$ has a private noisy observation $X^i_t$ of it at each time $t$. The private observations of players are conditionally independent given $V$. We then specialize this setting to the case of a Linear Quadratic Gaussian (LQG) non-zero-sum game where $V$ is a Gaussian random variable and players' observations are generated through a linear Gaussian model from $V$. Our LQG model closely follows that of~\cite{VaAn16a} with one important difference: the private observations of players in~\cite{VaAn16a} are independent where in our case, they are dependent through $V$; in particular they are conditionally independent given $V$.
	Our model can also be thought of as a generalization of the one in~\cite{BiHiWe92} where $V$ models the value of a product (or a technology) and agents receive a noisy private signal about it and decide whether to adopt it or not, with the important difference that we allow multiple agents to act simultaneously and, unlike~\cite{BiHiWe92}, we also allow them to return to the marketplace at each time instance and receive a new observation on $V$.
	
	One of the contributions of this paper is to show that, due to the conditional independence of the private signals given $V$, the private belief chain stops at the second step and players beliefs over others' beliefs are public functions of their own beliefs (the first step beliefs). In the LQG model, we further show that the beliefs are Gaussian and hence, are characterized by their mean and covariance matrix. Furthermore, the players estimation over others' estimations are public linear functions of their own estimations. We hypothesize (and eventually prove) structured PBE with strategies for user $i$ being linear in $\hat{V}^i_t$, the private estimate of $V$ by user $i$, generated by a (private) Kalman filter. This is the second contribution of this work.
	
	We show that the equilibrium strategies can be characterized by an appropriate backward sequential decomposition algorithm akin to dynamic programming. In the LQG model, the main difference of our work from the standard stochastic control LQG framework is that the forward recursion that evaluates covariance matrices cannot be performed separately as it depends on the equilibrium strategies. This was also the case in~\cite{VaAn16a}.
	A unique feature of our development is the requirement to update in a forward manner additional quantities that are observation dependent (public actions). This precludes off-line evaluation of these forward-updated quantities  and necessitates their inclusion as part of the state of the above mentioned backward sequential decomposition.
	This is the third contribution of this work.

	\subsection{Literature Review}
	
	In this section we give an overview of the related literature with a focus on the information sructures. In ~\cite{ho1972team}, a framework, called precedence diagram, was introduced to characterize the information structures in team problems with asymmetric information.  The evolving (dynamic) information of the decision makers is modeled by  a different (new) controller making a decision at each time with the specific information corresponding to that time available to her. The authors have also provided some examples of the dynamic team problems, one of which is LQG team problem with nested information structure and have proved optimality of  linear controllers. The specific information structure considered, nested information, allows the authors to form an equivalent static team problem for the dynamic model considered and hence, avoiding furthur challenges of dealing with dynamic models.
	
	LQG models have been studied extensively for decision and control problems. In the simplest instance of a single centralized controller it is well known that there is separation of estimation and control, posterior beliefs of the state are Gaussian, a sufficient statistic for control is the state estimate evaluated by the Kalman filter, the optimal control is linear in the state estimate, and the required covariance matrices can be calculated offline~\cite{KuVa86}. Although it is known that, in general, linear controllers are not optimal in LQG team problems~\cite{witsenhausen1968counterexample}, as we mentioned, some information structures have been identified for which linear controllers are shown to be optimal such as the works with nested information structure ~\cite{ho1972team}, stochastically nested information structure \cite{yuksel2009stochastic} and partial history sharing information structure  \cite{mahajan2015sufficient}. Private beliefs do not emerge in these models because of the specific information structure considered. In the nested information structure, there is no need to form beliefs to interpret the action of the predecessors because the decision maker already knows their information. In the model considered in \cite{mahajan2015sufficient}, the decision makers have local memory (not perfect memory) and the authors have not defined any summaries for the history and therefore, beliefs and hence, private beliefs are not introduced.

	In order to capture the strategic behavior of agents,
	dynamic decision problems have also been considered in the context of dynamic games and there is extensive literature on dynamic  games with asymmetric information. In \cite{basar1978two},  the author considers a delayed observation sharing model where all of the  previous private observations are shared with all of the players and the asymmetry of the information is only due to the private observations at current time. This specific information structure avoids the private beliefs in the sufficient statistics because they can be formed by augmenting the public belief by the current private observation. One-step delayed information sharing is also used in~\cite{altman2009stochastic}. Similarly, in~\cite{BiHiWe92,BiAn18,BiHeAn19j, HeBiAn19}, there is a public belief that can be augmented by the players' static private signals,  to form the private beliefs.
	

	%
	Authors in~\cite{GuNaLaBa14b} have used the common information approach, which breaks the history into the common and private parts and similarly, two partial strategies are introduced. One is applied to the private part of the history and the other one generates the first one based on the public part of the history. Finding the strategy that is generated  based on the public part of the history does not have the challenges of asymmetric games because the public part of the history is common between all players.  The solution concept used is called common information based Markov perfect equilibria.  Note that in~\cite{GuNaLaBa14b}, the private part of the history is not summarized into any other quantity, and therefore, no private beliefs had to be defined. A similar approach is used in~\cite{OuTaTe17}.
	
	In~\cite{VaAn16a}, authors have considered a multi-stage LQG game and characterized a signaling equilibrium which is linear in agents' private observations. In addition, a backward sequential decomposition was presented for the construction of the equilibrium, based on the general development in~\cite{vasal2018systematic}. In this work, the private observations are independent across agents and therefore there are no private beliefs in the game. This is  because a player's belief over others' private observations is independent of her private observation and hence, the belief is public.

	A number of works consider LQG games where information available to some players is affected by the decision of others.
	The works of~\cite{crawford1982strategic} on strategic information transmission, and~\cite{farokhi2014gaussian} on Gaussian cheap talk consider two-stage games and focus on Bayesian Nash equilibria. These works, however, consider games that are not dynamic. This implies that there is no need to search for the sufficient statistics and no private belief will be defined.
	The classic work on Bayesian persuasion~\cite{kamenica2011bayesian}, and the related one on strategic deception~\cite{sayin2018dynamic} consider two-stage and multi-stage games, respectively, and focus on (sender preferred) subgame perfect equilibria owing to the fact that strategies (as opposed to only the actions) of the sender are observed. Although the authors of~\cite{sayin2018dynamic} consider a dynamic game, they do not summarize the  history into time invariant quantities and they search for the strategies over the whole time horizon. Therefore, although the problem becomes intractable for large time horizons, the issue of private beliefs does not appear.
	
	The unique feature of this work is that we consider dependent private observations (specifically, conditionally independent on a hidden state of the world) between agents, in conjunctions with strategies with time-invariant domains, and so sufficient statistics (beliefs) are defined. As a result, we are forced to deal with private beliefs and the aforementioned issue of the infinite sequence of beliefs on beliefs has to be resolved. This is what makes the considered model interesting and more challenging compared to the previous works.

	The remaining part of the paper is structured as follows. In Section~\ref{model-g} the general model is described. Section~\ref{solcon} is a review of the solution concept that we have considered in this paper. We develop our main results in Section~\ref{sPBE}. In section~\ref{lqg}, we describe the special case of the model that is an LQG game, followed by the development of a concrete example in Section~\ref{sec:example} together with numerical results. We conclude in Section~\ref{conclusion}. Most of the proofs of theorems and lemmas are relegated to the Appendices at the end of the paper.

	\subsection{Notation}
	We use upper case letters for scalar and vector random variables and lower case letters for their realizations.
	We use the notation $\P(a|b)$ to denote the probability $\P(A=a|B=b)$ for discrete random variables and to denote the measure $\P(da|B=b)$ for continuous random variables.
	Bold upper case letters are used to denote matrices. Subscripts denote time indices and superscripts represent player identities. The notation $-i$ denotes the set of all players except $i$. All vectors are column vectors. The transpose of a matrix $\mat{A}$ (or vector) is denoted by $\mat{A}'$.  We use  semicolons $``;"$ for vertical concatenation of matrices (or vectors). For any vector (or matrix) with time and player indices, $a^i_t$ (or $\mat{A}^i_t$), $a^{-i}_t$ denotes the vertical concatenation of vectors (or matrices) $a^1_t,a^2_t,\hdots,a^{i-1}_t,a^{i+1}_t,\hdots$. Further, $a^i_{1:t}$ means  $(a^i_1,a^i_2,\hdots,a^i_t)$. In general, for any vector with time and player indices, $a^i_t$, we remove the superscript to show the vertical concatenation of the whole vectors and we remove the subscript to show the set of all vectors for all times.
	The matrix of all zeros with appropriate dimensions is denoted by $\mat{0}$ and the identity matrix of appropriate dimensions is denoted by $\mat{I}$. For two matrices $\mat{A}$ and $\mat{B}$, $\mathfrak{D}(\mat{A},\mat{B})$ represents the block diagonal concatenation of these matrices, i.e., $\left[\begin{array}{cc}
	\mat{A} & \mat{0} \\ \mat{0} & \mat{B}
	\end{array}\right]$ (it applies for any number of matrices). By $\mathfrak{D}(\mat{A}^{-i})$, we mean the block diagonal concatenation of matrices $\mat{A}^j$ for $j \in -i$. Further, $\qd(A;B)$ represents $B'AB$.
	For the equation $\left[\begin{array}{ccc}\tilde{a} \ ;& \tilde{b} \ ;&  \tilde{c}
	\end{array}\right]=\mat{A}\left[\begin{array}{ccc}a  \ ;&  b \ ;&  c
	\end{array}\right]$, the notation $(\mat{A})_{\tilde{a},b}$ denotes the submatrix of $\mat{A}$ corresponding to rows $\tilde{a}$ and columns $b$. We use $``:"$ for either of the row or column subscripts to indicate the whole rows or columns, e.g., $(\mat{A})_{:,b}$ denotes the submatrix of $\mat{A}$ corresponding to columns $b$. The trace of matrix $\mat{A}$ is denoted by $\tr(\mat{A})$.
	We use $\delta(\cdot)$ for the Dirac delta function. For any Euclidean set $\mathcal{S}$, $\Delta(\mathcal{S})$ represents the space of all probability measures on $\mathcal{S}$.

	\section{Model}\label{model-g}
	We consider a  discrete time dynamic system with $\cN=\{1,2,...,N\}$ strategic players over a finite time horizon $\cT=\{1,2,...,T\}$.   There is a static unknown state of the world $V\sim Q_V(\cdot)$. Each player has a private noisy observation $X^i_t$ of $V$ at every time step $t\in\cT$. At time $t$, player $i$ takes action $a^i_t \in \cA^i$ which is observed publicly by all players. The private observations are generated according to the kernel $X^i_t\sim Q_X^i(\cdot|V,A_{t-1})$ and they are independent across agents given $V$ and $A_{t-1}$, i.e.,
\begin{subequations}
	\begin{align}
	\P(X_t|V,A_{1:t-1},X_{1:t-1}) &=\P(X_t|V,A_{t-1}) \\
 &=\prod_{i \in \cN}Q_X^i(X^i_t|V,A_{t-1})
	\end{align}
\end{subequations}
	We assume that players have perfect recall and we can construct the history of the system at time $t$ as $h_t=(v,x_{1:t},a_{1:t-1}) \in \cH_t $ and the information set of player $i$ at time $t$ as $h^i_t=(x^i_{1:t},a_{1:t-1}) \in \cH^i_t$. At the end of time step $t$, each player $i$ receives the reward $r^i_t(v,a_t)$. We assume that the rewards are not observed by the players until the end of the time horizon.
	
	Let $g^i=(g^i_t)_{t \in \cT}$ be a probabilistic strategy of player $i$, where $g^i_t: \cH^i_t \rightarrow \Delta(\cA^i)$, meaning that player $i$'s action at time $t$ is generated according to the distribution $A^i_t\sim g^i_t(\cdot|h^i_t)$. The strategy profile of all players is denoted by $g$. For the strategy profile $g$, player $i$'s total expected reward is
	\begin{equation}
	J^{i,g}:=\E^g\left\{\sum_{t=1}^T r^i_t(V,A_t)\right\},
	\end{equation}	
	and her objective is to maximize her total expected reward.

	\section{Solution concept}\label{solcon}
	
	We can model this system as a dynamic game with asymmetric information and an appropriate solution concept for such games is Perfect Bayesian Equilibrium (PBE). A PBE consists of a pair $(\beta,\mu)$ (an assessment) of strategy profile $\beta=(\beta^i_t)_{t\in \cT , i \in \cN}$ and belief system  $\mu=(\mu^i_t)_{t\in \cT , i \in \cN}$ where $\mu^i_t : \cH^i_t \rightarrow \Delta(\cH_t)$ satisfies Bayesian updating and sequential rationality holds. Bayesian updating includes both on and off equilibrium histories.  This condition requires the beliefs to be Bayesian updated, if possible, given any history, whether that history is on equilibrium or off equilibrium~\cite{watson2017general}. To be more specific, given history $h^i_t$, which could be on or off-equlibrium, and the realizations at time $t$, i.e., $a_t,x^i_{t+1}$, the beliefs should be updated according to Bayes rule if $\mathbb{P}^g(a_t,x^i_{t+1}|h^i_t)>0$. Otherwise, the beliefs could be updated arbitrarily.
	For any $i \in \cN, \  t \in \cT, h^i_t \in \cH^i_t, \tilde{\beta}^i$, sequential rationality imposes the following condition for the strategy profile $\beta$:
	\begin{align}
	\E^{\beta^i\beta^{-i}}_{\mu}\left\{\sum_{n=t}^T r^i_n(V,A_n)|h^i_t\right\} \geq   \E^{\tilde{\beta}^i\beta^{-i}}_{\mu}\left\{\sum_{n=t}^T r^i_n(V,A_n)|h^i_t\right\}
	\end{align}

	\section{Structured PBE}\label{sPBE}
	The domain of the strategies $g^i_t(\cdot|h^i_t)$ is expanding in time. Finding such strategies is complicated with the complexity growing exponentially with the time horizon. For this reason, we consider summaries for $h^i_t \in \cH^i_t$, i.e., $S(h^i_t)$, that are time invariant. We are interested in PBEs with strategies, $g^i_t(\cdot|h^i_t)=\psi^i_t(\cdot|S(h^i_t))$, that are functions of $h^i_t$ only through the summaries $S(h^i_t)$. These PBEs are called structured PBEs~\cite{vasal2018systematic}. In contrast to $\cH^i_t$,  the set of summaries does not grow in time and therefore, finding such structured PBEs is less complicated than a general PBE.  According to \cite{vasal2018systematic}, we  can show that players can guarantee the same rewards by playing structured strategies compared to the general non-structured ones. In  dynamic games with asymmetric information, summaries are usually the belief of players over the unknown variables of the game.
	
	Define the private beliefs over the unknown state of the world $V$ as
	\begin{align}
	\xi^i_t(v)=\P^g(v|h^i_t)=\P^g(v|x^i_{1:t},a_{1:t-1})
	\end{align}
	We further define the conditional public belief over  the private beliefs as follows
	\begin{align}
	\pi_t(\xi_t|v)&=\P^g(\xi_t|v,h_t)=\P^g(\xi_t|v,a_{1:t-1}).
	\end{align}
		\begin{lemma}[Conditional Independence of Private Beliefs]\label{lm:conInd}
		We have the following equation for the  conditional public belief
		\begin{align}
		\pi_t(\xi_t|v)=\prod_{i\in \cN}\pi^i_t(\xi_t^i|v),
		\end{align}
		where $\pi^i_t(\xi_t^i|v)=\P(\xi_t^i|v,a_{1:t-1})$.
		Similarly, we have
		\begin{align}\label{eq:ind_x}
		\P^g(x_{1:t}|v,a_{1:t-1})=\prod_{i \in \cN}\P^g(x^i_{1:t}|v,a_{1:t-1}).
		\end{align}
	\end{lemma}
	
	\begin{IEEEproof}
		See Appendix \ref{prf:conInd}.
	\end{IEEEproof}
	Note that this conditional independence holds regardless of the strategy profiles $g$. Using this result, and with a slight abuse of notation\footnote{We will be using $\pi_t$ to denote the joint conditional $\pi_t(\xi_t|v)$ as well as the vector of marginal conditionals $\pi_t=[\pi^1_t, \ldots, \pi^N_t]$. The distinction will be obvious from the context.}, we can summarize  the conditional public belief into the vector $\pi_t=[\pi^1_t, \ldots, \pi^N_t]$.
	
	We are interested in strategies of the form $A^i_t \sim \psi^i_t(\cdot|\xi^i_t,\pi_t)=\gamma^i_t(\cdot|\xi^i_t)$, where $\gamma^i_t=\theta^i_t(\pi_t)$ and we will prove that such structured strategies form a PBE of the game.
	Note that with the above decomposition of the strategy $\psi$ into partial strategies $\gamma$ and the strategy $\theta$,
	designing  strategies $\psi$ is equivalent to designing  $\theta$.
	
	\subsection{Belief Update}
	In this subsection, we present two lemmas regarding the beliefs and their update rules.
	

\begin{lemma}\label{lm:blfupdt}
The private beliefs can be updated as $\xi^i_{t+1}=F^i(\xi^i_t,\pi^{-i}_t,\gamma^{-i}_t,a_t,x^i_{t+1})$, where $F^i$ is defined through
\begin{align}
\xi^i_{t+1}(v)
 &=\frac{\begin{multlined}\int_{\xi^{-i}_t}\xi^i_t(v)\prod_{j\in-i}\pi^{j}_t(\xi^{j}_t|v)\gamma^j_t(a_t^j|\xi_t^j)Q^i_X(x^i_{t+1}|v,a_t)\end{multlined}}
        {\begin{multlined}\int_{\xi^{-i}_t,\tilde{v}}\xi^{i}(\tilde{v})\prod_{j\in-i}\pi^j_t(\xi^j_t|\tilde{v})\gamma^j_t(a_t^j|\xi_t^j)Q^i_X(x^i_{t+1}|\tilde{v},a_t)\end{multlined}}.
\end{align}
\end{lemma}
\begin{IEEEproof}
		See Appendix \ref{prf:blfupdt}.
\end{IEEEproof}
Note that this update depends on the strategy profile $g$ only through the partial function $\gamma_t^{-i}$, i.e., it is independent of the strategy $\theta$.
We will also use the notation $\xi_{t+1}=F(\xi_t,\pi_t,\gamma_t,a_t,x_{t+1})$ for the update function of the vector of private beliefs.

	\begin{lemma}\label{lm:pubblfupdt}
		The conditional public beliefs can be updated as $\pi^i_{t+1}=F^i_{\pi}(\pi_t,\gamma_t,a_t)$, where $F^i_{\pi}$ is defined through
		 	\begin{align}
\pi^i_{t+1}(\xi^i_{t+1}|v)=\frac{\begin{multlined}\int_{\xi^i_t,x^i_{t+1}}\pi^i_t(\xi^i_{t}|v)\gamma^i_t(a^i_t|\xi_t^i) Q_X^i(x^i_{t+1}|v,a_t)\\\textbf{1}_{F^i(\xi^i_t,\pi^{-i}_t,\gamma^{-i}_t,a_t,x^i_{t+1})}(\xi^i_{t+1}) \end{multlined}}{\begin{multlined} \int_{\tilde{\xi}^i_t}\pi^i_t(\tilde{\xi}^i_t)\gamma^i_t(a^i_t|\tilde{\xi}_t^i)\end{multlined}}.
		 \end{align}
	\end{lemma}
	\begin{IEEEproof}
		See Appendix \ref{prf:pubblfupdt}.
	\end{IEEEproof}
	Similar to the previous lemma, this update depends on the strategy profile $g$ only through the partial function $\gamma_t$, i.e., it is independent of the strategy $\theta$. We use the notation $\pi_{t+1}=F_{\pi}(\pi_t,\gamma_t,a_t)$ to denote the update function of the vector of conditional public beliefs.

	\subsection{Equilibrium Strategies}
	In this subsection, we will show that structured strategies of the form $\gamma^i_t(\cdot|\xi^i_t)$, where $\gamma^i_t=\theta^i_t(\pi_t)$  form sPBE of the game. The following theorem formalizes this result and presents the fixed point equation characterizing the equilibrium strategies.
	
	\begin{theorem}\label{thm:mdp-g}
		The strategy profile $\gamma^*_t=\theta_t(\pi_t)$ characterized by the following fixed point equation, forms a sPBE of the game. For all $i \in \cN$,
		\begin{subequations}\label{spbe}
			\begin{align}
			&\gamma^{*,i}_t(\cdot|\xi^i_t)=\arg \max_{\gamma^i_t(\cdot|\xi^i_t)}
			\E   [ \hat{r}^i_t(\pi_t,\xi^i_t,A^i_t) \nonumber \\
			&\ +J^i_{t+1}(F_{\pi}(\pi_t,\gamma^*_t,A_t),F^i(\xi^i_t,\pi_t,\gamma^{*,-i}_t,A_t,X^i_{t+1})))| \pi_t,\xi^i_t], \label{spbe1}\\
			&J^i_{t}(\pi_t,\xi^i_t)=\max_{\gamma^i_t(\cdot|\xi^i_t)}
			\E   [ \hat{r}^i_t(\pi_t,\xi^i_t,A^i_t) \nonumber \\
			&\ +J^i_{t+1}(F_{\pi}(\pi_t,\gamma^*_t,A_t),F^i(\xi^i_t,\pi_t,\gamma^{*,-i}_t,A_t,X^i_{t+1})))| \pi_t,\xi^i_t],
			\end{align}
			where, $\hat{r}^i_t(\pi_t,\xi^i_t,a^i_t)=\E\left[r^i_t(V,A_t)|\pi_t,\xi^i_t,a^i_t\right]$.
		\end{subequations}
	\end{theorem}
	
	\begin{IEEEproof}
		See Appendix~\ref{prf:mdp-g}.
	\end{IEEEproof}
	We remark here that in equation \eqref{spbe} the update rule of the public belief $\pi_t$ is using the equilibrium strategies $\gamma^*_t$  and therefore, for each time instance $t$, the collection of equations of the form \eqref{spbe1} for all $i\in\cN$ constitutes a fixed point equation over the strategy profile $\gamma^*_t$.
The reason for this is that in characterizing a PBE, one needs to fix the belief structure and then finds the equilibrium strategies corresponding to those beliefs. On the other hand, the beliefs have to be consistent with the equilibrium strategies. This creates a fixed point equation over $\gamma^{*,i}_t$. Furthermore, the above equation has to be solved simultaneously for all $i \in \cN$, thus creating the fixed point equation over the strategy $\gamma^{*}_t$.

	\subsection{Discussion}
	In this section, we characterized the sufficient statistics of the histories of the considered dynamic game. As we mentioned in the Introduction, these summaries include private beliefs, $\xi^i_t$.
	One may wonder how we resolved the issue with the chain of private beliefs that was discussed in the Introduction.
	In other words, how did we resolve the issue of possibly requiring an infinite hierarchy of beliefs on beliefs.
	In the previous development, we actually proved that this chain stops at the second step.
	To see this, consider the introduction of private beliefs over others' private beliefs, i.e., $\P(\xi^{-i}_t|h^i_t)$.
	The results of Lemma~\ref{lm:conInd} show that
	\begin{subequations}
		\begin{align}
		\P&(\xi^{-i}_t|h^i_t) \nonumber \\
		&= \int_v \P(\xi^{-i}_t,v|h^i_t) \\
		&= \int_v \P(\xi^{-i}_t|v,h^i_t) \P(v|h^i_t) \\
		&= \int_{v,x^{-i}_{1:t}} \P(\xi^{-i}_t|v,h^i_t,x^{-i}_{1:t}) \P(x^{-i}_{1:t}|v,h^i_t) \P(v|h^i_t) \\
		&\stackrel{(a)}{=} \int_{v,x^{-i}_{1:t}} \P(\xi^{-i}_t|a_{1:t-1},x^{-i}_{1:t}) \P(x^{-i}_{1:t}|v,a_{1:t-1}) \P(v|h^i_t) \\
		&= \int_v \P(\xi^{-i}_t|v,a_{1:t-1}) \P(v|h^i_t) \\
		&= \int_v \pi_t(\xi^{-i}_t|v) \xi^i_t(v),
		\end{align}
	\end{subequations}
	where (a) is due to the definition of the private beliefs and~\eqref{eq:ind_x}.
	The above implies that these beliefs can be evaluated by the public information, $\pi_t$, and the first order private beliefs $\xi^i_t$.  This is the exact reason why $\pi_t(\xi_t|v)$ was defined.

	\section{LQG Model} \label{lqg}
	In this section, we study a specific instance of the model discussed so far which is the case where the unknown state of the world, $V$, is a Gaussian random variable,  the  private observation kernels are linear and Gaussian and the instantaneous  reward  is quadratic. Therefore we have an LQG model. The motivation for studying this model stems from the general development in the previous section. In particular we required that equilibrium strategies are generated based on private beliefs and public beliefs on beliefs. In the LQG setting these beliefs can be greatly simplified, thus enabling us to more succinctly characterize  the equilibrium strategies discussed in the previous section.
	
	In this model, we consider an unknown state of the world $V\sim N(\vc{0},\mat{\Sigma})$ with size $N_v$. Each player has a private noisy observation $X^i_t$ of $V$ at every time step $t\in\cT$
	\begin{equation}
	x^i_t=v+w^i_t,
	\label{pr-ob-lqg}
	\end{equation}
	where $W^i_t \sim N(\vc{0},\mat{Q}^i)$ and all of the noise random vectors $W^i_t$ are independent across $i$ and $t$ and also independent of $V$. The values of $\mat{\Sigma}$ and $\mat{Q}^i, \ \forall i \in \cN$ are common knowledge between players. Note that in order to maintain the linearity of private observations, we have considered uncontrolled private observations unlike the general model in first part of the paper.  More discussion on this matter can be found in section \ref{extensions}. We have $a^i_t \in \cA^i=\mathbb{R}^{N_a}$. The instantaneous reward\footnote{Unlike more standard LQG setting we consider ``rewards'' instead of ``costs'' to maintain consistency with the general problem discussed earlier.} is given by
	\begin{equation}
	r^i_t(v,a_t)
	= \left[{\begin{array}{cc}v' & a_t' \end{array}}\right]
	\mat{R}^i_t
	\left[\begin{array}{c} v \\ a_t \end{array}\right]
	= \qd(\mat{R}^i_t;\left[\begin{array}{c} v \\ a_t \end{array}\right] ),
	\end{equation}
	where $\mat{R}^i_t$ is a symmetric matrix of appropriate dimensions.

	\subsection{Equilibrium Beliefs}
	
	In this setting, we will show that  the private beliefs $\xi^i_t$ are Gaussian and since any Gaussian belief can be expressed in terms of its mean and covariance matrix, we define the summaries such that they include the mean and covariance matrices of the beliefs of the players over $V$. The mean of each player's belief, i.e., her estimate of $V$, will be her private information. The covariance matrix, however, can be calculated publicly. We define the private estimate of players over $V$ as follows. For all $ i \in \cN, \ t \in \cT$,
	\begin{align}
	\hv^i_t&=\E[V| h^i_t]=\E[V|x^i_{1:t},a_{1:t-1}],
	\end{align}
	Since the private beliefs can be expressed in terms of their means and covariance matrices and since the covariance matrices are publicly calculated, the conditional public belief $\pi_t^i(\xi^i_t|v)$ is equivalent to a belief over the private estimates. Intuitively, each player, in addition to her own estimate of $V$, needs to interpret actions of others and predict their future actions. Hence, each player needs to have a belief over the estimates of other players on $V$. We will show that this latter belief is also Gaussian and therefore, one needs to keep track of only its mean and covariance.   We define the following quantity for all $ i \in \cN, \ t \in \cT$,
	\begin{align}
	\hhv^{i,j}_t&=\E[\chv^j_t|h^i_t]=\E[\chv^j_t|x^i_{1:t},a_{1:t-1}].
	\end{align}	
	The quantity $\hv^i_t$ is player $i$'s best estimate of $V$ given her observations up to time $t$. As mentioned before, this quantity is a private estimation for player $i$ and is not measurable with respect to the sigma algebra generated by the observations of any other player $j$. Hence, player $i$ should form an estimate over the private estimates of other players and this is the reason $\hhv^{i,j}_t$ is defined. This in turn implies that players' strategies should also be a function of their estimates over others' estimates of $V$. Hence, the same argument as the one in the first part of the paper about private beliefs holds and we need to define an estimate over estimates of players over other players' estimates of $V$. This argument continues as long as these estimates are private. Therefore, once again, we are faced with the problem of having to define a chain of private beliefs which are expressed as private estimates in this model. This chain stops whenever one of the estimates of players is public (or a public function of previous-step private estimates) and therefore, there is no need to form an estimate over it.
	
	Indeed, we will show that $\hhv^{i,-i}_t$ is a public linear function of $\hv^i_t$, hence,  there is no need to include $\hhv^{i,-i}_t$ in the private part of the summary $S(h^i_t)$ and therefore, no other player needs to form an estimate over it. The summary we use for $h^i_t$ is defined as $S(h^i_t)=(\hv^i_t, P(h^i_t))$, where $P(h^i_t)$ is the public summary for $h^i_t$ and it includes the covariance matrix of player $i$'s belief over $V$ and some other needed quantities that will be subsequently defined.
	We are interested in equilibria with strategies of the form $A^i_t \sim \psi^i_t(\cdot|\hv^i_t,P(h^i_t))=\gamma^i_t(\cdot|\hv^i_t)$, where $\gamma^i_t=\theta^i_t(P(h^i_t))$. In particular, we want to prove that pure linear strategies of the form $\gamma^i_t(a^i_t|\hv^i_t)=\delta(a^i_t-\mat{L}^i_t\hv^i_t-\vc{m}^i_t)$, where $\mat{L}^i_t$ and $\vc{m}^i_t$ are matrices with appropriate dimensions and are functions of $P(h^i_t)$, form a PBE of the game.

In the next theorem, we show that when linear strategies are employed, the private beliefs are Gaussian.
	\begin{theorem}\label{thm:lqg-belief}
Assuming pure linear strategies of the form $\gamma^i_t(a^i_t|\hv^i_t)=\delta(a^i_t-\mat{L}^i_t\hv^i_t-\vc{m}^i_t)$,
$\forall t\in\cT$ and  $\forall i \in \cN$, the private belief $\xi^i_t$ on $V$ is Gaussian $N(\hv^i_t,\Sigma^i_t)$, where $\hv^i_t$ is the private estimate of player $i$ of $V$ and $\Sigma^i_t$ is the corresponding covariance matrix, which can be evaluated publicly.
Consequently, the public belief $\pi^i_t(\xi^i_t|v)$ can be reduced to a belief $\pi^i_t(\hv^i_t|v)$.
Furthermore, $\pi^i_t(\hv^i_t|v)$ is Gaussian with mean  $\mat{E}^i_tv+\vc{f}^i_t$, where matrices $\mat{E}^{i}_t$, $\vc{f}^i_t$ can be evaluated publicly.


	\end{theorem}
	\begin{IEEEproof}
		See Appendix~\ref{prf:gm}.
	\end{IEEEproof}
	
	In the following we summarize the parameters needed to update each of  the quantities introduced in the proof of Theorem~\ref{thm:lqg-belief} and we introduce update functions for each one.
\begin{subequations}
	\begin{align}
\hv^i_{t+1}&=F_{\hv}(\hv^i_t,\mat{\Sigma}^i_{t+1|t},\mat{E}^{-i}_t,\vc{f}^{-i}_t,\mat{L}^{-i}_t,\vc{m}^{-i}_t,a^{-i}_t,x^i_{t+1})\label{vhfun}\\
\mat{\Sigma}^i_{t+1}&=F_{\mat{\Sigma}^i}(\mat{\Sigma}^i_{t+1|t},\mat{L}^{-i}_t)\label{sigmafun}\\
\mat{\Sigma}_{t+2|t+1}&=F_{\mat{\Sigma}}(\mat{\Sigma}_{t+1|t},\mat{E}_t,\mat{L}_t)\\
\tilde{\mat{\Sigma}}_{t+2|t+1}&=F_{\tilde{\mat{\Sigma}}}(\tilde{\mat{\Sigma}}_{t+1|t},\mat{\Sigma}_{t+1|t},\mat{E}_t,\mat{L}_t)\\
\mat{E}_{t+1}&=F_{\mat{E}}(\mat{E}_t,\mat{\Sigma}_{t+1|t},\tilde{\mat{\Sigma}}_{t+1|t},\mat{L}_t)\\
\vc{f}_{t+1}&=F_{\vc{f}}(\vc{f}_t,\mat{\Sigma}_{t+1|t},\tilde{\mat{\Sigma}}_{t+1|t},\mat{E}_t,\mat{L}_t,\vc{m}_t,a_t)
\end{align}
\label{sumsum}
\end{subequations}
Equations \eqref{vhfun} and \eqref{sigmafun} correspond to the private belief update and are similar in structure to the update function $F^i$ of of $\xi^i_t$ in Lemma~\ref{lm:blfupdt} for the general case. The remaining update functions correspond to the public belief update $F_{\pi}$ in Lemma~\ref{lm:pubblfupdt} for the general case.

Note that according to the above equations, the quantities $\mat{\Sigma}_{t+1|t}$, $\tilde{\mat{\Sigma}}_{t+1|t}$, $\mat{E}_t$ are updated recursively using the strategy matrices $\mat{L}_t$. Hence, if one  knows the strategies, one can calculate these quantities offline for the entire time horizon of the game. However, the quantity $\vc{f}_k$ is updated using the strategy matrices $\mat{L}_t$ and vectors $m_k$ as well as the realized actions $a_t$ and therefore, they cannot be evaluated offline.

We reiterate at this point that  Theorem~\ref{thm:lqg-belief} implies that the estimate of player $i$ over private estimates of players $-i$, i.e., $\hhv^{i,-i}_t$, is a linear function of $\hv^i_t$,
	\begin{subequations}
		\begin{align}
		\hhv^{i,-i}_t
		&= \E[\hat{V}^{-i}_t|h^i_t] \\
		&= \E[\E[\hat{V}^{-i}_t|V,A_{1:t-1}]|h^i_t] \\
		&= \E[\mat{E}^{-i}_t V +\vc{f}^{-i}_t |h^i_t] \\
		&= \mat{E}^{-i}_t \hv^i_t+\vc{f}^{-i}_t,
		\label{vhhupdate}
		\end{align}
	\end{subequations}
with matrices $\mat{E}^{-i}_t$ and $\vc{f}^{-i}_t$ being public information.
As a result, assuming linear strategies of the form $a^i_t=\mat{L}^i_t\hv^i_t+\vc{m}^i_t$ at equilibrium, one can form the summary  $S(h^i_t)=(\hat{v}^{i}_t,P(h^i_t))$ and base the selection of the matrices $\mat{L}^i_t$ and $\vc{m}^i_t$ on the public part of this summary, $P(h^i_t)$. In the next section we show that indeed linear strategies can form an equilibrium and provide a methodology to find the quantities $\mat{L}^i_t$ and $\vc{m}^i_t$.

	\subsection{Linear Structured PBE}
	Theorem \ref{thm:lqg-belief} implies that $S^i_t=\left[\begin{array}{cc} V \ ; & \chv^{-i}_{t-1} \end{array}\right]$ is a jointly Gaussian random vector conditioned on player $i$'s observation till time $t$, $\forall i \in \cN, t \in \cT$. This implies that the beliefs over $V$ are jointly Gaussian and so players need only keep track of their belief's mean (estimation) and covariance matrices. Furthermore, this theorem implies that a player's belief over other players beliefs is also Gaussian and hence, players need to keep track of their estimation on other players' estimations, i.e., $\hhv$. The important point of Theorem~\ref{thm:lqg-belief} is the statement that the estimation of players on others' estimations is a linear function of their own estimation and hence, in order to keep track of the estimation over other players' estimations, a player only needs to keep track of her own estimation over $V$. Therefore,  $\hv^i_t$ is a sufficient statistic for player $i$'s private observations till time $t$.
	
In terms of the public summary,  we see four public quantities, $\mat{\Sigma}_{t+1|t}$, $\tilde{\mat{\Sigma}}_{t+1|t}$, $\mat{E}_t$  and $\vc{f}_t$ in \eqref{sumsum}.  With some abuse of notation, we define $\mat{\Sigma}_t=[\mat{\Sigma}_{t+1|t}, \tilde{\mat{\Sigma}}_{t+1|t}]$. We will show that the tupple $(\mat{\Sigma}_t, \mat{E}_t,\vc{f}_t)$ is the public summary of $h^i_t$, i.e., $P(h^i_t)$. Note that $\mat{E}_t$  and $\vc{f}_t$ are involved in the expression for the mean of the conditional public belief over $\hv_t$, hence, they correspond to the conditional public belief $\pi_t$ in the first part of the paper. The convariance matrices $\mat{\Sigma}_{t+1|t}$, $\tilde{\mat{\Sigma}}_{t+1|t}$ represent the covariance matrices of the private and conditional public beliefs. This implies that by having the tuple $(\hv^i_t,\mat{\Sigma}_t, \mat{E}_t,\vc{f}_t)$,  we have full characterization of the private and public belief and therefore, we have the summaries for the LQG game.
	
	 Therefore, we consider strategies of the form $\psi^i_t(\cdot|\hv^i_t,\mat{\Sigma}_t,\mat{E}_t,\vc{f}_t)=\gamma^i_t(\cdot|\hv^i_t)$.	In particular, we will now show that linear strategies of the form $\gamma^i_t(\cdot|\hv^i_t)=\delta(a^i_t-\mat{L}^i_t\hv^i_t-m^i_t)$, where  $\mat{L}_t$ and $m_t$ are derived from $(\mat{\Sigma}_t, \mat{E}_t,\vc{f}_t)$, are PBE of the game.

%
	\begin{theorem}
		The strategy profile $\psi^i_t(\cdot|\hv^i_t,\mat{\Sigma}_t,\mat{E}_t,\vc{f}_t)=\gamma^i_t(\cdot|\hv^i_t) \ \forall i \in \cN$ where $\gamma^i_t(\cdot|\hv^i_t)=\delta(a^i_t-\mat{L}^i_t\hv^i_t-m^i_t)$, together with the corresponding Gaussian beliefs derived in Theorem~\ref{thm:lqg-belief}, form a structured PBE of the game.
		
		The strategy matrices $\mat{L}_t$ and vectors $\vc{m}_t$ are constructed throughout the proof.
		\label{thm:linstlqg}
	\end{theorem}
	
	\begin{IEEEproof}
		See Appendix \ref{prf:linstlqg}.
	\end{IEEEproof}		
	
	 One important result from the proof of Theorem \ref{thm:linstlqg} is that the reward to go, $J^i_{t}(\hv^i_t,\mat{\Sigma}_t,\mat{E}_t,\vc{f}_t)$ is quadratic with respect to $\hv^i_t$ and $\vc{f}_t$, which are the only quantities in the summary that can not be evaluated offline, i.e., we have
	 \begin{align}
J^i_t(\hv^i_t,\mat{\Sigma}_t,\mat{E}_t,\vc{f}_t)=\qd(\mat{Z}^i_t;\left[\begin{array}{c} \hv^i_t\\\vc{f}_t \end{array}\right])+\vc{z}^{i\prime}_t\left[\begin{array}{c} \hv^i_t\\\vc{f}_t \end{array}\right]+o^i_t.
	 \end{align}
Therefore, if we have the quantities $\mat{Z}^i_t$, $\vc{z}^{i\prime}_t$, and $o^i_t$	we can evaluate the reward to go for every value of $\hv^i_t$ and $\vc{f}_t$.

In the following, we propose a backward algorithm that evaluates the quantities $\mat{Z}^i_t$, $\vc{z}^{i}_t$, and $o^i_t$ as well as the strategy matrices $\mat{L}_t$, $\mat{M}_t$  and vectors $\bar{m}_t$ (we have $m^i_t=\mat{M}^i_t\vc{f}_t+\bar{m}^i_t$, according to the proof of Theorem \ref{thm:linstlqg}) as functions of  $(\mat{\Sigma}_t, \mat{E}_t)$.
Before stating the algorithm, we define the following functions.
	\begin{subequations}
     	\begin{align}
         \mat{L}_t&=g_{\mat{L},t}(\mat{\Sigma}_t, \mat{E}_t)\\
         \mat{M}_t&=g_{\mat{M},t}(\mat{\Sigma}_t, \mat{E}_t)\\
         \bar{m}_t&=g_{\bar{m},t}(\mat{\Sigma}_t, \mat{E}_t)\\
         \mat{Z}_t&=\psi_{\mat{Z},t}(\mat{\Sigma}_t, \mat{E}_t)\\
         z_t &=\psi_{z,t}(\mat{\Sigma}_t, \mat{E}_t)\\
         o_t&=\psi_{o,t}(\mat{\Sigma}_t, \mat{E}_t)
     \end{align}
     \label{backfun}
	\end{subequations}

	 \textbf{Backward Algorithm (Offline)}
	\begin{enumerate}
		\item Set $t=T$. Set $\mat{Z}_{T+1}=\psi_{\mat{Z},T+1}(\mat{\Sigma}_{T+1}, \mat{E}_{T+1})=\mat{0}$, $z_{T+1}=\psi_{z,T+1}(\mat{\Sigma}_{T+1}, \mat{E}_{T+1})=\mat{0}$ and $o_{T+1}=\psi_{\mat{Z},T+1}(\mat{\Sigma}_{T+1}, \mat{E}_{T+1})=\mat{0}$ for every $\mat{\Sigma}_{T+1},\mat{E}_{T+1}$.
		\item Calculate $\mat{L}_t=g_{\mat{L},t}(\mat{\Sigma}_t, \mat{E}_t)$, $\mat{M}_t=g_{\mat{M},t}(\mat{\Sigma}_t, \mat{E}_t)$,  $\bar{m}_t=g_{\bar{m},t}(\mat{\Sigma}_t, \mat{E}_t)$, and  $\mat{Z}_t=\psi_{\mat{Z},t}(\mat{\Sigma}_t, \mat{E}_t)$ for every $\mat{\Sigma}_t,\mat{E}_t$ and the corresponding $\psi_{\mat{Z},t+1}(\cdot, \cdot)$ according to equation \eqref{strategy} and \eqref{Zt}.\\
		\item Set $t=t-1$.
		\item If $t\geq 1$ Go to step 3. Else stop.
	\end{enumerate}

Using the functions defined above, one can run the following forward algorithm to find the strategy matrices $\mat{L}_t$,   $\mat{M}_t$  and vectors $\bar{m}_t$ and the quantities $\mat{Z}^i_t$, $\vc{z}^{i\prime}_t$, and $o^i_t$.

	 \textbf{Forward Algorithm (Offline)}
	\begin{enumerate}
		\item Set $t=1$.
		\item Initialize the value of $\mat{\Sigma}_1$ and $\mat{E}_1$ according to equations \eqref{sigma} and \eqref{sigmatilde}.
		\item Using $\mat{\Sigma}_t$ and $\mat{E}_t$, find  $\mat{L}_t$, $\mat{M}_t$, $\bar{m}_t$ and the quantities $\mat{Z}^i_t$, $\vc{z}^{i\prime}_t$, and $o^i_t$  according to equation \eqref{backfun}.
		\item Using $\mat{\Sigma}_t$, $\mat{E}_t$ and $\mat{L}_t$, calculate $\mat{\Sigma}_{t+1}$ and $\mat{E}_{t+1}$ according to equations \eqref{sumsum}.
		\item Set $t=t+1$.
		\item If $t\leq T$, Go to step 3. Else stop.
	\end{enumerate}

	\subsection{Model Extensions}\label{extensions}
	In this section, we investigate  alternative models that can be studied with the methodology introduced in this paper and we explain how the results can be extended to such models.
	
	As it is clear in equation \eqref{pr-ob-lqg}, in the LQG model considered in this paper, the private observations are not controlled by the actions, unlike the general model of the first part of the paper. If we were to add control actions to equation \eqref{pr-ob-lqg}, in order to maintain linearity, we would have added a term such as $B^i_ta_t$ and therefore, equation \eqref{pr-ob-lqg} would have looked like $x^i_t=v+w^i_t+B^i_ta_t$. Since the actions are publicly observed, the amount of information that player $i$ extracts from $V$ remains the same with or without the term $B^i_ta_t$. Hence, because the private observations serve only as measurements of $V$, adding control to equation \eqref{pr-ob-lqg} does not make any difference in the results.
	
	Controlled private observations could make a difference in the LQG model if the private observations could affect the instantaneous rewards. That is, if the reward was $r^i_t(v,a_t,x^i_t)= \qd(\mat{R}^i_t;\left[\begin{array}{c} v \\ a_t \\x^i_t\end{array}\right])$. Note that the amount of information that $x^i_t$ conveys about $V$ is still the same as in the uncontrolled case. We can show that results similar to all of the ones in this paper will hold for this model with controlled private observations and this type of instantaneous reward.  Note that in this case, the strategies woud be linear in both the private estimation and the latest private observation.
	
	We can also extend our results of the first part of the paper (the general model) to a model with the instantaneous reward being of the form of $r^i_t(v,a_t,x^i_t)$. In this case, $x^i_t$  should be added to the summaries and the results will hold.
	
			\section{Example}\label{sec:example}
	In this section, we describe some numerical examples to show the equilibrium strategies discussed in this paper. In these examples, we derive the equilibrium strategies by solving a fixed point equation for the entire time horizon using the following algorithm. Note that the superscript $(k)$ in $A^{(k)}$ denotes the number of iterations performed. We define the convergence error as $\epsilon^{(k)}=\max(|\mat{L}_{1:T}^{(k+1)}-\mat{L}_{1:T}^{(k)}|, |\mat{M}_{1:T}^{(k+1)}-\mat{M}_{1:T}^{(k)}|, |\bar{m}_{1:T}^{(k+1)}-\bar{m}_{1:T}^{(k)}|)$.

\textbf{Numerical Algorithm (Offline)}
	\begin{enumerate}
		\item Set $k=1$.
		\item Initialize $\mat{L}_{1:T}^{(1)}$, $\mat{M}_{1:T}^{(1)}$, and $\bar{m}_{1:T}^{(1)}$ arbitrarily.
		\item Using $\mat{L}_{1:T}^{(k)}$, evaluate $\mat{\Sigma}_{1:T}^{(k+1)}$, $\mat{E}_{1:T}^{(k+1)}$ according to equations \eqref{sumsum} in a forward manner (using initial conditions $\mat{\Sigma}_1$ and $\mat{E}_1$ according to equations \eqref{sigma} and \eqref{sigmatilde}).
		\item Using $\mat{L}_{1:T}^{(k)}$, $\mat{M}_{1:T}^{(k)}$, $\bar{m}_{1:T}^{(k)}$,
                and $\mat{\Sigma}_{1:T}^{(k+1)}$, $\mat{E}_{1:T}^{(k+1)}$,
                evaluate $\mat{L}_{1:T}^{(k+1)}$, $\mat{M}_{1:T}^{(k+1)}$, and $\bar{m}_{1:T}^{(k+1)}$ according to the backward algorithm.
		\item Evaluate $\epsilon^{(k)}$. If it is below the desired threshold, stop. Otherwise, go to step 4.
	\end{enumerate}
	Note that in each step of the backward algorithm, one needs to solve a fixed point equation with respect to the strategy matrices and vectors to derive the functions defined in eq. \eqref{backfun} (see eq. \eqref{strategy} in Appendix \ref{prf:linstlqg}). However, in the numerical algorithm described above, we use the last iteration quantities for the right hand side of the equations and consequently, we do not need to solve any fixed point equations.

As a concrete example, we consider a setting where there is a project with an unknown attribute denoted by $v$. There are two agents working on this project exerting a costly effort $a^i_t$. The agents are rewarded based on the alignment of their effort with the project attribute, $v$, as well as based on their cooperation. At each time slot, the agents have private observations, $x^i_t$, of the project attribute. We consider two instances of the game where $v$ is scalar in one and a two dimensional vector in the other, while the efforts are scalars in both.

	\subsection{Scalar State and Action}

We model the considered scenario for scalar $v$ and scalar actions $a^i_t$ with the  instantaneous rewards being $R^1_t(v,a_t)=a^1_tv+\frac{1}{2}a^1_ta^2_t-{(a^1_t)}^2$ and $R^2_t(v,a_t)=a^2_tv+\frac{1}{2}a^1_ta^2_t-{(a^2_t)}^2$. That is, we set $\mat{R}^1_t=\left[\begin{array}{ccc} 0 & \frac{1}{2} & 0 \\ \frac{1}{2} &  -1 & \frac{1}{4}\\ 0 & \frac{1}{4} & 0 \end{array}\right]$ and $\mat{R}^2_t=\left[\begin{array}{ccc} 0 & 0 & \frac{1}{2} \\ 0 &  0 & \frac{1}{4}\\ \frac{1}{2}& \frac{1}{4} & -1\end{array}\right]$. Note that the term $a^i_tv$ in the instantaneous rewards accounts for the alignment of $a^i_t$ with $v$, and the term $a_t^1a_t^2$ denotes the cooperation between the agents.

Case~1:
If we assume that agents perfectly observe $V$, i.e., if we set $\mat{Q}^1=0$ and $\mat{Q}^2=0$, the following linear equilibrium strategy matrices and vectors are derived from the numerical analysis of this game for $T=2$ and $\mat{\Sigma}=1$
\begin{subequations}
\begin{align}
\begin{array}{cc}
\mat{L}^1_1=\frac{2}{3}  \quad &\mat{L}^2_1=\frac{2}{3} \\
\mat{L}^1_2=\frac{2}{3}  \quad  &\mat{L}^2_2=\frac{2}{3} .
\end{array}
\end{align}
%
%
\end{subequations}
Furthermore, we have  $\bar{m}^i_t=0$ for $t=1,2$ and $i=1,2$. Note that since in this case, $f_t=0$ for $t=1,2$, the strategy matrices $\mat{M}^i_t$ will not play any roles and are not presented here.  These results imply that each agent will exert effort exactly equal to $\frac{2}{3} V$. As it turns out, these strategies are myopic, i.e., we also observe these strategies in the case $T=1$.  The reason for having myopic strategies  is that the observations are perfect and hence, the actions have no effect in shaping the future beliefs.

Case~2: Consider agents with equally imperfect observations, $\mat{Q}^1=\mat{Q}^2=1$. The following strategy matrices are derived
\begin{subequations}
\begin{align}
\begin{array}{cc}
\mat{L}^1_1=0.6722 \quad & \mat{L}^2_1=0.6722\\
\mat{L}^1_2=0.5333 \quad  &\mat{L}^2_2=0.5333
\end{array}
\end{align}
\begin{align}
\begin{array}{cc}
\hspace{-0.3cm} \mat{M}^1_1=\left[0.0561  \quad  0.2620\right]  \quad & \hspace{-0.2cm} \mat{M}^2_1=\left[0.2620   \quad  0.0561\right]\\
\hspace{-0.3cm} \mat{M}^1_2=\left[0.0356   \quad  0.1422\right] \quad  & \hspace{-0.2cm} \mat{M}^2_2=\left[0.1422   \quad  0.0356\right],
\end{array}
\end{align}
%
\end{subequations}
together with  $\bar{m}^i_t=0$ for $t=1,2$ and $i=1,2$.
Once more, it is observed that $\bar{m}^i_t=0$ and as will be seen, the same is happening  in all of the other cases studied as well.  This could imply that it is sufficient to restrict attention to strategies with zero $\bar{m}^i_t$. We also observe that the value of the strategy matrices decrease with time.

Case~3: If one agent has better observations than the other, i.e., $\mat{Q}^1=1$, $\mat{Q}^2=2$, the strategy matrices are changed as follows.
\begin{subequations}
	\begin{align}
	\begin{array}{cc}
	\mat{L}^1_1=0.6700 \quad & \mat{L}^2_1=0.6619\\
	\mat{L}^1_2=0.5224 \quad  &\mat{L}^2_2=0.5373
	\end{array}
	\end{align}
	\begin{align}
	\begin{array}{cc}
	\hspace{-0.3cm} \mat{M}^1_1=\left[0.0520   \quad  0.2701\right]  \quad & \hspace{-0.1cm} \mat{M}^2_1=\left[0.2738 \quad   0.0605\right]\\
	\hspace{-0.3cm} \mat{M}^1_2=\left[0.0348  \quad  0.1433\right] \quad  & \hspace{-0.1cm} \mat{M}^2_2=\left[0.1393 \quad   0.0358\right]
	\end{array}
	\end{align}
%
\end{subequations}
and $\bar{m}^i_t=0$ for $t=1,2$ and $i=1,2$. One can explain these results by  paying attention to the interactions between the agents. At $t=1$, agent one has a better estimation of $V$ compared to agent two and therefore, she has higher $\mat{L}^1_1$. At $t=2$, agent two has learned the estimation of agent one through her action at $t=1$ and therefore, the two agents have almost equal estimations. But this time, agent two exerts slightly higher effort to compensate agent one's efforts at $t=1$.

Case~4: The interaction between agents can also be seen in a scenario  where one agent has perfect observations and the other one has partial observations, i.e.,  $\mat{Q}^1=0$, $\mat{Q}^2=2$. The strategy matrices are given as follows.
\begin{subequations}
	\begin{align}
	\begin{array}{cc}
	\mat{L}^1_1=0.7125 \quad & \mat{L}^2_1= 0.6781\\
	\mat{L}^1_2=0.5000 \quad  &\mat{L}^2_2=0.6250
	\end{array}
	\end{align}
	\begin{align}
	\begin{array}{cc}
	\hspace{-0.3cm} \mat{M}^1_1=\left[0.0142 \quad   0.1808\right]  \quad & \hspace{-0.2cm} \mat{M}^2_1=\left[0.1817  \quad   0.0452\right]\\
	\hspace{-0.3cm} \mat{M}^1_2=\left[0.0333 \quad   0.1667\right] \quad  & \hspace{-0.2cm} \mat{M}^2_2=\left[0.1333  \quad   0.0417\right],
	\end{array}
	\end{align}
\end{subequations}
and $\bar{m}^i_t=0$ for $t=1,2$ and $i=1,2$.

Case~5: Finally, consider a case where both agents have very noisy observations, that is $\mat{Q}^1$, $\mat{Q}^2$ are large numbers.  In this case, $\hv^i_t=0$ and $f_t=0$. Therefore, the strategy matrices $\mat{L}^i_t$ and $\mat{M}^i_t$ do not play any roles and the actions will only follow $\bar{m}^i_t$. For this game we obtain $\bar{m}^i_t=0$ for $t=1,2$ and $i=1,2$.
%

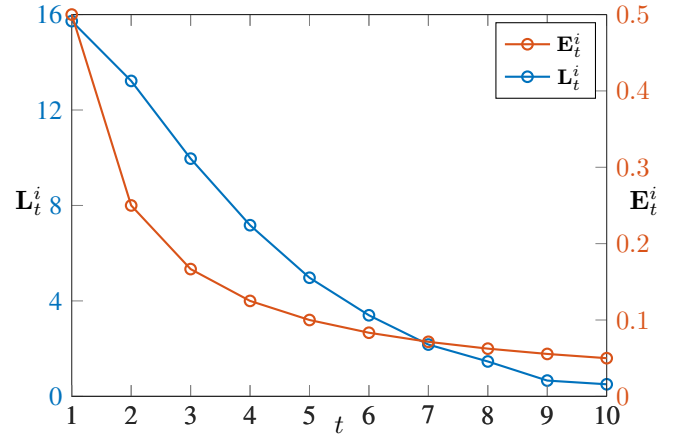
\begin{figure}[h]
\centering
\begin{tikzpicture}

\definecolor{mycolor1}{rgb}{0.15,0.15,0.15}
\definecolor{mycolor2}{rgb}{0,0.447,0.741}
	\definecolor{mycolor3}{rgb}{0.85,0.325,0.098}

\begin{axis}[
axis y line*=left,
x label style={at={(axis description cs:0.5,-0.03)},anchor=north},
y label style={at={(axis description cs:-0.08,0.45)},rotate=270,anchor=south},
xlabel={$t$},
ylabel={$\mat{L}^i_t$},
scale only axis,
every outer x axis line/.append style={mycolor1},
every x tick label/.append style={ font=\color{mycolor1}},
every outer y axis line/.append style={mycolor1},
every y tick label/.append style={font=\color{mycolor2}},
width=2.8in,
height=2in,
xmin=1, xmax=10,
ymin=0, ymax=16,
xtick={1,2,3,4,5,6,7,8,9,10},
xticklabels={1,2,3,4,5,6,7,8,9,10},
ytick={0,4,8,12,16},
yticklabels={0,4,8,12,16},
axis on top]

\addplot [
color=mycolor2,
solid,
mark=o,
line width=0.8pt
]
coordinates{
	(1,15.7249)
	(2,13.2185)
	(3,9.96259)
	(4,7.17234)
	(5,4.97115)
	(6,3.39813)
	(7,2.17068)
	(8,1.46262)
	(9,0.662135)
	(10,0.506329)

}; \label{Lit}

\end{axis}
\begin{axis}[
scale only axis,
axis y line*=right,
x label style={at={(axis description cs:0.5,0.03)},anchor=north},
y label style={at={(axis description cs:1.07,0.45)},rotate=270,anchor=south},
ylabel={$\mat{E}^i_t$},
every outer x axis line/.append style={mycolor1},
every x tick label/.append style={ font=\color{mycolor1}},
every outer y axis line/.append style={mycolor1},
every y tick label/.append style={font=\color{mycolor3}},
width=2.8in,
height=2in,
xmin=1, xmax=10,
ymin=0, ymax=0.5,
xtick={1,2,3,4,5,6,7,8,9,10},
xticklabels={1,2,3,4,5,6,7,8,9,10}]
\addplot [
color=mycolor3,
solid,
mark=o,
line width=0.8pt
]
coordinates{
	(1,0.5)
	(2,0.25)
	(3,0.166667)
	(4,0.125)
	(5,0.1)
	(6,0.0833333)
	(7,0.0714286)
	(8,0.0625)
	(9,0.0555556)
	(10,0.05)
}; \label{Eit}
\addlegendimage{/pgfplots/refstyle=Lit}\addlegendentry{\footnotesize{$\mat{E}^i_t$}}
\addlegendimage{/pgfplots/refstyle=Lit}\addlegendentry{\footnotesize{$\mat{L}^i_t$}}
\end{axis}
\end{tikzpicture}
\caption{ \small Strategy matrices $\mat{L}^i_t$ and quantities $\mat{E}^i_t$ for $T=10$.}
\label{strategyT10}
\end{figure}

Case~6: We have also derived the strategy matrices of the game  for larger values of $T$.  In Figure \ref{strategyT10}, we can see the plot of the strategy matrices $\mat{L}^i_t$ with respect to time for the symmetric case of $\mat{Q}^1=\mat{Q}^2=1$ and for $T=10$. As before, we observe a trend where as time goes by, the values of the strategy matrices decrease. The intuition behind why such behavior is observed is that more public information is observed as time goes by. Therefore, the players estimation over others' estimations is mainly characterized by the public part of the state, $f_t$, rather than the private estimates. This indicates that the matrix $\mat{E}_t$ decreases with time and as it is obvserved in our numerical results in Figure \ref{strategyT10}, it converges to zero. One can also see that the strategies decrease as $\mat{E}_t$ decreases. Therefore, the strategy matrices $\mat{L}_t$ decrease as time passes and they converge to $0.5$, which is the equilibrium of the game  when $\mat{E}_t=\mat{0}$.

\subsection{Game vs Centralized LQG}
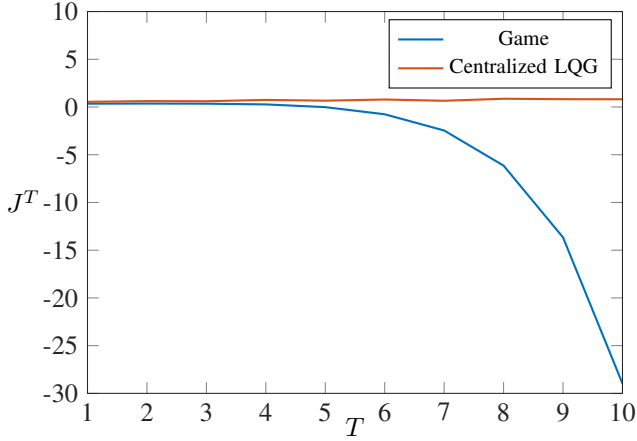
\begin{figure}[h]
	\centering	
	\begin{tikzpicture}
	\definecolor{mycolor1}{rgb}{0.15,0.15,0.15}
	\definecolor{mycolor2}{rgb}{0,0.447,0.741}
	\definecolor{mycolor3}{rgb}{0.85,0.325,0.098}
	
	\begin{axis}[
	x label style={at={(axis description cs:0.5,-0.04)},anchor=north},
	y label style={at={(axis description cs:-0.12,0.45)},rotate=270,anchor=south},
	xlabel={$T$},
	ylabel={$J^T$},
	scale only axis,
	every outer x axis line/.append style={mycolor1},
	every x tick label/.append style={font=\color{mycolor1}},
	every outer y axis line/.append style={mycolor1},
	every y tick label/.append style={font=\color{mycolor1}},
	width=2.8in,
	height=2in,
	xmin=1, xmax=10,
	ymin=-30, ymax=10,
	xtick={1,2,3,4,5,6,7,8,9,10},
	xticklabels={1,2,3,4,5,6,7,8,9,10},
		ytick={-30,-25,-20,-15,-10,-5,0,5,10},
	yticklabels={-30,-25,-20,-15,-10,-5,0,5,10},
	axis on top]
	
	\addplot [
	color=mycolor2,
	solid,
	line width=0.8pt
	]
	coordinates{
		(1,0.3493)
		(2,0.3623)
		(3,0.3485)
		(4,0.2724)
		(5,-0.013)
		(6,-0.762)
		(7,-2.4594)
		(8,-6.1414)
		(9,-13.6627)
		(10,-28.9737)

	};
	\addplot [
	color=mycolor3,
	solid,
		line width=0.8pt
	]
	coordinates{
		(1,0.543799)
		(2,0.621163)
		(3,0.60604)
		(4,0.742906)
		(5,0.6631)
		(6,0.783865)
		(7,0.65293)
		(8,0.866004)
		(9,0.819831)
		(10,0.81001)
		
	};
\legend{\footnotesize{Game}\\\footnotesize{Centralized LQG}\\}
	\end{axis}
	
	\end{tikzpicture}
	\caption{ \small Total rewards per time obtained in game vs centralized LQG. }
	\label{rewtogo}
\end{figure}

In this subsection, we have compared the total rewards per time obtained through the game by players for $\mat{Q}^1=\mat{Q}^2=1$ with a scenario in which both actions are taken by a single decision maker and the sum of the two rewards are collected by her. We have done this comparison for different time horizons $T$ and Figure \ref{rewtogo} depicts the plot of the total rewards per time obtained, $J^T$, in the two considered scenarios.
We notice that players are doing worse compared to the centralized decision maker, specifically as the time horizon increases. The reason is that in the game scenario, the uncertainty in predicting the average reward-to-go increases drastically as time horizon increases. The centralized decision maker, however, benefits from time horizon increasing and her total reward per time converges to one. The reason is that as time goes by, the estimation over  $V$  becomes better and better and the reward converges to the one in the complete information case.

\subsection{Two Dimensional State and Scalar Action}
In this part, we consider a two dimensional attribute vector for the project, i.e., $V$ is a two dimensional vector. Each agent tries to be aligned with one element of the attribute vector while maitaining the cooperation with the other agent. We can model this alignment and cooperation of agents with $R^1_t(v,a_t)=a^1_t v(1)+a^1_ta^2_t-{(a^1_t)}^2$ and $R^2_t(v,a_t)=a^2_tv(2)+a^1_ta^2_t-{(a^2_t)}^2$. That is, we set $\mat{R}^1_t=\left[\begin{array}{cccc} 0 & 0 & \frac{1}{2} & 0 \\ 0 & 0 & 0 & 0\\ \frac{1}{2} & 0 &  -1 & \frac{1}{2}\\ 0 & 0& \frac{1}{2} & 0 \end{array}\right]$ and $\mat{R}^2_t=\left[\begin{array}{cccc} 0 & 0 & 0 &  0\\  0 & 0 & 0 &  \frac{1}{2} \\ 0 &  0 & 0&  \frac{1}{2}\\ 0 & \frac{1}{2} & \frac{1}{2}& -1\end{array}\right]$.  We also set $\mat{\Sigma}=\left[\begin{array}{cc} 1 & 0\\0 & 1\end{array}\right]$.

Case~1: The following linear equilibrium strategy matrices are derived for the full information case.
\begin{subequations}
	\begin{align}
	\begin{array}{cc}
	\mat{L}^1_1=\left[\frac{2}{3} \quad  \frac{1}{3}\right] \quad  &  \mat{L}^2_1=\left[\frac{1}{3}  \quad  \frac{2}{3} \right] \\
	\mat{L}^1_2=\left[\frac{2}{3} \quad    \frac{1}{3}\right]  \quad  &\mat{L}^2_2=\left[\frac{1}{3}  \quad  \frac{2}{3} \right],
	\end{array}
	\end{align}
\end{subequations}
 and $\bar{m}^i_t=0$ for $t=1,2$ and $i=1,2$. Also, similar to the scalar case, $\mat{M}^i_t$ strategy matrices do not play any roles here since $f_t=0$. We see that if $V$ is perfectly observed, each agent will align her effort with a weighted average of $V(1)$ and $V(2)$ with the element correponding to that agent having twice the weight.  Also, similar to the scalar case, myopic strategies are played.

Case~2: Consider the partial information scenario with  $\mat{Q}^1=\left[\begin{array}{cc} 1 & 0\\0 & 1\end{array}\right]$ and $\mat{Q}^2=\left[\begin{array}{cc} 1 & 0\\0 & 1\end{array}\right]$. The following linear equilibrium strategy matrices are derived.
\begin{subequations}
	\begin{align}
\begin{array}{cc}
\mat{L}^1_1=\left[0.7224  \quad   0.2402\right] \quad  & \hspace{-0.05cm} \mat{L}^2_1=\left[0.2402   \quad  0.7224 \right] \\
\mat{L}^1_2=\left[0.4858  \quad   0.0842\right]  \quad  & \hspace{-0.05cm} \mat{L}^2_2=\left[0.0842 \quad   0.4858\right]
\end{array}
	\end{align}
	\begin{align}
	\mat{M}^1_1&=\left[0.2874  \quad   0.0780  \quad   0.1793   \quad  0.6054\right] \\
	 \mat{M}^2_1& =\left[0.6054  \quad  0.1793  \quad  0.0780  \quad  0.2874\right]\\
	\mat{M}^1_2&=\left[0.1619 \quad    0.0281  \quad   0.0561 \quad    0.3239\right]\\
	 \mat{M}^2_2&=\left[0.3239  \quad  0.0561  \quad  0.0281  \quad  0.1619\right],
	\end{align}
\end{subequations}
and $\bar{m}^i_t=0$ for $t=1,2$ and $i=1,2$. Similar to the scalar scenario, we observe that the value of the strategy matrices decrease with time and again, $\bar{m}^i_t=0$ for all of the cases.

Case~3: If each agent fully observes her corresponding element of the state and partially observes the other one, i.e., $\mat{Q}^1=\left[\begin{array}{cc} 0 & 0\\0 & 1\end{array}\right]$ and $\mat{Q}^2=\left[\begin{array}{cc} 1 & 0\\0 & 0\end{array}\right]$, we have the following linear equilibrium strategy matrices.
\begin{subequations}
	\begin{align}
	\begin{array}{cc}
	\mat{L}^1_1=\left[0.7198  \quad  0.4232\right] \quad  & \hspace{-0.05cm} \mat{L}^2_1=\left[0.4232   \quad  0.7198 \right] \\
	\mat{L}^1_2=\left[0.5071   \quad  0.2055\right]  \quad  & \hspace{-0.05cm} \mat{L}^2_2=\left[0.2055  \quad  0.5071\right]
	\end{array}
	\end{align}
	\begin{align}
	\mat{M}^1_1&=\left[0.3196   \quad  0.1506  \quad   0.3235   \quad  0.6293\right] \\
	\mat{M}^2_1& =\left[0.6293  \quad  0.3235  \quad  0.1506  \quad  0.3196\right]\\
	\mat{M}^1_2&=\left[0.1690  \quad   0.0685  \quad   0.1370  \quad  0.3380\right]\\
	\mat{M}^2_2&=\left[0.3380  \quad  0.1370   \quad  0.0685   \quad  0.1690\right]
	\end{align}
\end{subequations}
and $\bar{m}^i_t=0$ for $t=1,2$ and $i=1,2$. An intuitive reason of why the second element and the first element of the strategy matrices $\mat{L}^1_t$ and $\mat{L}^2_t$, respectively, are larger than the previous case is that the second element and the first element of $\mat{E}^1_t$ and $\mat{E}^2_t$, respectively, have increased.
	\section{Conclusion} \label{conclusion}
	In this paper, we studied  a dynamic game with asymmetric information and dependent types and we characterized the structured perfect Bayesian equilibria of the game. We also studied a special case of our model that was Linear Quadratic Gaussian (LQG) non-zero-sum game and we characterized linear structured perfect Bayesian equilibria for the game. One of the important points that we made in this paper was that due to the conditional independence of the private signals, the private belief chain stops at the second step and players beliefs over others' beliefs are public functions of their own beliefs. We further proved that these beliefs are Gaussian in the LQG  case.

	A future direction for this research could be investigating the models for which we have the same interesting features for the beliefs as we do in this paper. That is, the models for which the private belief chain stops at two or any other given number of steps.

	\appendices{}
	
	\section{Proof of Lemma~\ref{lm:conInd}}
	\label{prf:conInd}
%
	\begin{subequations}
\begin{align}
&\pi_t(\xi_t|v)  =\P(\xi_t|v,a_{1:t-1}) \\
&=\frac{\int_{x_{1:t}}\P(x_{1:t},\xi_t,a_{1:t-1}|v)}{\int_{x_{1:t}}\P(x_{1:t},a_{1:t-1}|v)}
\\
&=\frac{ \hspace{-0.2cm} \begin{multlined}\int_{x_{1:t}} \prod_{i\in \cN} \prod_{s=1}^{t-1} Q_X^i(x^i_{s}|v,a_{s-1})\P(a_s^i|x^i_{1:s},a_{1:s-1})\\ Q_X^i(x_t^i|v,a_{t-1})  \P(\xi_t^i|x^i_{1:t},a_{1:t-1}) \end{multlined}}{\begin{multlined}\int_{x_{1:t}}  \prod_{i\in \cN} \prod_{s=1}^{t-1} Q_X^i(x^i_{s}|v,a_{s-1})\P(a_s^i|x^i_{1:s},a_{1:s-1})\\Q_X^i(x_t^i|v,a_{t-1})\end{multlined}}
\\
&=\prod_{i\in \cN}\hspace{-0.14cm} \frac{\hspace{-0.2cm}  \begin{multlined} \int_{x^i_{1:t}} \prod_{s=1}^{t-1} Q_X^i(x^i_{s}|v,a_{s-1})\P(a_s^i|x^i_{1:s},a_{1:s-1})\\ Q_X^i(x_t^i|v,a_{t-1}) \P(\xi_t^i|x^i_{1:t},a_{1:t-1}) \end{multlined}}{\begin{multlined}\int_{x^i_{1:t}} \prod_{s=1}^{t-1} Q_X^i(x^i_{s}|v,a_{s-1})\P(a_s^i|x^i_{1:s},a_{1:s-1})\\Q_X^i(x_t^i|v,a_{t-1})\end{multlined}}
\\
&=\prod_{i\in \cN}\frac{\P(\xi^i_t,a_{1:t-1}|v)}{\P(a_{1:t-1}|v)}=\prod_{i\in \cN}\P(\xi^i_t|v,a_{1:t-1})\\&=\prod_{i\in \cN}\pi_t(\xi^i_t|v).
\end{align}
	\end{subequations}

	The second part of the theorem is similarly proved as follows.
	\begin{subequations}
	\begin{align}
&\P(x_{1:t}|v,a_{1:t-1}) =\frac{\P(x_{1:t},a_{1:t-1}|v)}{\P(a_{1:t-1}|v)} \\
&=\frac{  \begin{multlined} \prod_{i\in \cN} \prod_{s=1}^{t-1} Q_X^i(x^i_{s}|v,a_{s-1})\P(a_s^i|x^i_{1:s},a_{1:s-1})\\Q_X^i(x_t^i|v,a_{t-1}) \end{multlined}}
{  \begin{multlined} \int_{x_{1:t}}\prod_{i\in \cN} \prod_{s=1}^{t-1} Q_X^i(x^i_{s}|v,a_{s-1})\P(a_s^i|x^i_{1:s},a_{1:s-1})\\Q_X^i(x_t^i|v,a_{t-1}) \end{multlined}} \\
&=\frac{ \begin{multlined} \prod_{i\in \cN} \prod_{s=1}^{t-1} Q_X^i(x^i_{s}|v,a_{s-1})\P(a_s^i|x^i_{1:s},a_{1:s-1})\\Q_X^i(x_t^i|v,a_{t-1}) \end{multlined}}
{ \begin{multlined}\prod_{i\in \cN} \int_{x^i_{1:t}} \prod_{s=1}^{t-1} Q_X^i(x^i_{s}|v,a_{s-1})\P(a_s^i|x^i_{1:s},a_{1:s-1})\\Q_X^i(x_t^i|v,a_{t-1}) \end{multlined}} \\
&= \prod_{i\in \cN}\frac{ \begin{multlined} \prod_{s=1}^{t-1} Q_X^i(x^i_{s}|v,a_{s-1})\P(a_s^i|x^i_{1:s},a_{1:s-1})\\Q_X^i(x_t^i|v,a_{t-1}) \end{multlined}}
{ \begin{multlined} \int_{x^i_{1:t}} \prod_{s=1}^{t-1} Q_X^i(x^i_{s}|v,a_{s-1})\P(a_s^i|x^i_{1:s},a_{1:s-1})\\Q_X^i(x_t^i|v,a_{t-1}) \end{multlined}} \\
&=\prod_{i\in \cN} \frac{  \P(x^i_{1:t},a_{1:t-1}|v)}{ \P(a_{1:t-1}|v)} \\
&=\prod_{i\in \cN} \P(x^i_{1:t}|a_{1:t-1},v).
\end{align}
	\end{subequations}

	\section{Proof of Lemma~\ref{lm:blfupdt}}\label{prf:blfupdt}
	
	Using Bayes rule we have
	\begin{subequations}
		\begin{align}
		&	\xi^i_{t+1}(v) \nonumber \\
 &=\P(v|x^i_{1:t+1},a_{1:t})\\&=\frac{\P(v,x^i_{t+1},a_t|x^i_{1:t},a_{1:t-1})}{\P(x^i_{t+1},a_t|x^i_{1:t},a_{1:t-1})}\\
 &=\frac{\int_{\xi^{-i}_t}\P(v,x^i_{t+1},a_t,\xi^{-i}_t|x^i_{1:t},a_{1:t-1})}
        {\int_{\xi^{-i}_t,\tilde{v}}\P(\tilde{v},x^i_{t+1},a_t,\xi^{-i}_t|x^i_{1:t},a_{1:t-1})}\\
 &=\frac{\begin{multlined}\int_{\xi^{-i}_t} \P(v|x^i_{1:t},a_{1:t-1})\P(\xi^{-i}_t|v,a_{1:t-1})\\\P(a_t|\xi^{-i}_t,v,x^i_{1:t},a_{1:t-1})Q^i_X(x^i_{t+1}|v,a_t)\end{multlined}}
        {\begin{multlined}\int_{\xi^{-i}_t,\tilde{v}} \P(\tilde{v}|x^i_{1:t},a_{1:t-1})\P(\xi^{-i}_t|\tilde{v},a_{1:t-1})\\\P(a_t|\xi^{-i}_t,\tilde{v},x^i_{1:t},a_{1:t-1})Q^i_X(x^i_{t+1}|\tilde{v},a_t)\end{multlined}}\\
 &=\frac{\begin{multlined}\int_{\xi^{-i}_t} \xi^i_t(v)\pi^{-i}_t(\xi^{-i}_t|v)\prod_{j\in \cN}\gamma^j_t(a_t^j|\xi_t^j)Q^i_X(x^i_{t+1}|v,a_t)\end{multlined}}{\begin{multlined}\int_{\xi^{-i}_t,\tilde{v}} \xi^i_t(\tilde{v})\pi^{-i}_t(\xi^{-i}_t|\tilde{v})\prod_{j\in\cN}\gamma^j_t(a_t^j|\xi_t^j)Q^i_X(x^i_{t+1}|\tilde{v},a_t)\end{multlined}}\\
&=\frac{\begin{multlined}\int_{\xi^{-i}_t}\xi^i_t(v)\prod_{j\in-i}\pi^{j}_t(\xi^{j}_t|v)\gamma^j_t(a_t^j|\xi_t^j)Q^i_X(x^i_{t+1}|v,a_t)\end{multlined}}
        {\begin{multlined}\int_{\xi^{-i}_t,\tilde{v}}\xi^{i}(\tilde{v})\prod_{j\in-i}\pi^j_t(\xi^j_t|\tilde{v})\gamma^j_t(a_t^j|\xi_t^j)Q^i_X(x^i_{t+1}|\tilde{v},a_t)\end{multlined}}.
		\end{align}
	\end{subequations}

	\section{Proof of Lemma~\ref{lm:pubblfupdt}}\label{prf:pubblfupdt}
	
	Using Bayes rule we have
	\begin{subequations}
		\begin{align}
		&\pi^i_{t+1}(\xi^i_{t+1}|v) \nonumber \\
		&=\P(\xi^i_{t+1}|v,a_{1:t})\\
		&=\frac{\int_{\xi_t,x^i_{t+1}}\P(\xi^i_{t+1},\xi_t,x^i_{t+1},a_t|v,a_{1:t-1})}{\int_{\xi_t}\P(\xi_t,a_t|v,a_{1:t-1})}\\
		&=\frac{\begin{multlined}\int_{\xi_t,x^i_{t+1}}\P(\xi_{t}|v,a_{1:t-1})\P(a_t|\xi_t,a_{1:t-1})\P(x^i_{t+1}|v,a_t)\\\textbf{1}_{F^i(\xi^i_t,\pi^{-i}_t,\gamma^{-i}_t,a_t,x^i_{t+1})}(\xi^i_{t+1}) \end{multlined}}{\begin{multlined}\int_{\xi_t} \P(\xi_t|v,a_{1:t-1})\P(a_t|\xi_t,a_{1:t-1})\end{multlined}}
		\\
		&=\frac{\begin{multlined}\int_{\xi_t,x^i_{t+1}}\prod_{j \in \cN}\pi^j_t(\xi^j_{t}|v)\gamma^j_t(a^j_t|\xi_t^j)Q_X^i(x^i_{t+1}|v,a_t)\\\textbf{1}_{F^i(\xi^i_t,\pi^{-i}_t,\gamma^{-i}_t,a_t,x^i_{t+1})}(\xi^i_{t+1}) \end{multlined}}{\begin{multlined}\int_{\xi_t} \prod_{j\in \cN}\pi^j_t(\xi^j_t)\gamma^j_t(a^j_t|\xi_t^j)\end{multlined}}.\\
		&=\frac{\begin{multlined}\prod_{j \neq i}\int_{\xi^j_t} \pi^j_t(\xi^j_{t}|v)\gamma^j_t(a^j_t|\xi_t^j)\\ \int_{\xi^i_t,x^i_{t+1}}\pi^i_t(\xi^i_{t}|v)\gamma^i_t(a^i_t|\xi_t^i) Q_X^i(x^i_{t+1}|v,a_t)\\\textbf{1}_{F^i(\xi^i_t,\pi^{-i}_t,\gamma^{-i}_t,a_t,x^i_{t+1})}(\xi^i_{t+1}) \end{multlined}}{\begin{multlined} \prod_{j\in \cN}\int_{\xi^j_t}\pi^j_t(\xi^j_t)\gamma^j_t(a^j_t|\xi_t^j)\end{multlined}}\\&
		=\frac{\begin{multlined}\int_{\xi^i_t,x^i_{t+1}}\pi^i_t(\xi^i_{t}|v)\gamma^i_t(a^i_t|\xi_t^i) Q_X^i(x^i_{t+1}|v,a_t)\\\textbf{1}_{F^i(\xi^i_t,\pi^{-i}_t,\gamma^{-i}_t,a_t,x^i_{t+1})}(\xi^i_{t+1}) \end{multlined}}{\begin{multlined} \int_{\xi^i_t}\pi^i_t(\xi^i_t)\gamma^i_t(a^i_t|\xi_t^i)\end{multlined}}.
		\end{align}
	\end{subequations}

	\section{Proof of Theorem~\ref{thm:mdp-g}}\label{prf:mdp-g}
	To prove the theorem, we show that if every player $-i$ plays according to strategy $\gamma^{*,-i}_t=\theta^{-i}_t(\pi_t)$, the best response of player $i$ is of the form $\gamma^{*,i}_t=\theta^{i}_t(\pi_t)$ and it is derived from the given fixed point equation.
	We show that if we fix the update rule of $\pi_t$ to $\pi_{t+1}=F_{\pi}(\pi_t,\gamma^*_t,a_t)=F_{\pi}(\pi_t,\theta_t(\pi_t),a_t)$ and assume that player $i$ is forced to use these beliefs as her true beliefs, then she faces an MDP with state $(\pi_t, \xi^i_t)$,  action  $a^i_t$ and instantaneous reward  $\hat{r}^i_t(\pi_t,\xi^i_t,a^i_t)=\E\left[r^i_t(V,A_t)|\pi_t,\xi^i_t,a^i_t\right]$.

	We first need to prove that the state $(\pi_t,\xi^i_t)$ evolves according to a controlled Markov process. Indeed,
	\begin{align}
	\P&(\pi_{t+1},\xi^i_{t+1}|\pi_{1:t},\xi^i_{1:t},a^i_{1:t})= \nonumber \\
	&\int_{v,\xi^{-i}_t,a^{-i}_t,x^i_{t+1}} \pi^{-i}_t(\xi^{-i}_t|v)\xi^i_t(v)\theta^{-i}_t(\pi_t)(a^{-i}_t|\xi^{-i}_t)Q(x^i_{t+1}|v,a_t) \nonumber \\
	&\qquad \textbf{1}_{F_{\pi}(\pi_t,\theta_t(\pi_t),a_t)}(\pi_{t+1})
	\textbf{1}_{F^i(\xi^i_t,\pi^{-i}_t,\theta^{-i}_t(\pi_t),a_t,x^i_{t+1})}(\xi^i_{t+1})  \nonumber\\
	&=\P(\pi_{t+1},\xi^i_{t+1}|\pi_{t},\xi^i_{t},a^i_{t}).
	\label{eq:markov_property}
	\end{align}

	The average instantaneous reward can now be written as $\E[r^i_t(V,A_t)]= \E[ \E[r^i_t(V,A_t)|\Pi_t,\Xi^i_t,A^i_t]]$, where
	\begin{align}
	\E&[r^i(V,A_t)|\pi_t,\xi^i_t,a^i_t] \nonumber \\
	&=\int_{v,a^{-i}_t} r^i(v,a_t) \int_{\xi^{-i}_t}\P(v,a^{-i}_t,\xi^{-i}_t|\pi_t,\xi^i_t,a^i_t)  \nonumber \\
	&=\int_{v,a^{-i}_t} r^i(v,a_t) \int_{\xi^{-i}_t}\theta^{-i}_t(\pi_t)(a^{-i}_t|\xi^{-i}_t) \pi^{-i}_t(\xi^{-i}_t|v)\xi^i_t(v)  \nonumber \\
	&=\int_{v,a^{-i}_t} r^i(v,a_t) \int_{\xi^{-i}_t} \theta^{-i}_t(\pi_t)(a^{-i}_t|\xi^{-i}_t) \pi^{-i}_t(\xi^{-i}_t|v)\xi^i_t(v)  \nonumber \\
	&=: \hat{r}^i_t(\pi_t,\xi^i_t,a^i_t).
	\end{align}
	Based on the above, it is now clear that user $i$ faces an MDP and her best response strategy is the solution of the following backward dynamic program
	\begin{subequations}
		\begin{align}
		J^i_{t}(\pi_{t},\xi^i_{t})
		= &\max_{\gamma^i_t(\cdot|\xi^i_{t})} \E[ \hat{r}^i_t(\pi_t,\xi^i_t,a^i_t) +  J^i_{t+1}(\Pi_{t+1},\Xi^i_{t+1})| \pi_t,\xi^i_t,a^i_t] \\
		a^{*,i}_t\sim
		\arg &\max_{\gamma^i_t(\cdot|\xi^i_{t})} \E[\hat{r}^i_t(\pi_t,\xi^i_t,a^i_t) +  J^i_{t+1}(\Pi_{t+1},\Xi^i_{t+1})| \pi_t,\xi^i_t,a^i_t],
		\end{align}
		\label{MDP}
	\end{subequations}
	where expectation is wrt $\gamma^i_t$ and the conditional distribution in~\eqref{eq:markov_property}. Consequently the best response of user $i$ is of the form $A^{*,i}_t  \sim \psi^i_t(\cdot|\xi^i_t,\pi_t)$. Note that in the standard MDP formulation, it is suficient to only consider the pure strategies. However, in equation \eqref{MDP}, we  see randomized strategies. The reason of this modification is that in a PBE, the beliefs have to be consistent with the equilibrium strategies and we need $\psi^i_t(\cdot|\xi^i_t,\pi_t)=\gamma^{*,i}_t(\cdot|\xi^i_t)=\theta^i_t(\pi_t)(\cdot|\xi^i_t)$. Hence, the best responses satisfy the following fixed point equation at each time $t$. For all $i$ and all $\xi^i_t$ we have
	\begin{align}
	&\gamma^{*,i}(\cdot|\xi^i_t)=\arg \max_{\gamma^i(\cdot|\xi^i_t)}
	\E   [ \hat{r}^i_t(\pi_t,\xi^i_t,A^i_t) \nonumber \\
	&+J^i_{t+1}(F_{\pi}(\pi_t,\gamma^*_t,A_t),F^i(\xi^i_t,\pi_t,\gamma^{*,-i}_t,A_t,X^i_{t+1})))| \pi_t,\xi^i_t],
	\label{MDP-FP}
	\end{align}
	where expectation is wrt the distribution
	\begin{subequations}
		\begin{align}
	&	\P(a_t,x^i_{t+1}|\pi_{t},\xi^i_{t}) =\nonumber \\
		&\int_{\xi^{-i}_t,v} \gamma^i_t(a^i_t|\xi^i_t) \gamma^{*,-i}_t(a^{-i}_t|\xi^{-i}_t) \pi^{-i}_t(\xi^{-i}_t|v)\xi^i_t(v)Q_X^i (x^i_{t+1}|v,a_t).
		\end{align}
	\end{subequations}
	The above fixed point might not have a solution in pure strategies and therefore, we had to consider randomized strategies in equation \eqref{MDP}.

	\section{Proof of Theorem~\ref{thm:lqg-belief}}\label{prf:gm}
	Throughout this proof, the submatrices that are not explicitely specified are all zero matrices with appropriate dimensions.
	
	In order to prove the theorem we will define a dynamical system from the viewpoint of a specific user $i$ and show inductively that
	it is a Gauss Markov model. Gaussianity of both private and conditional public beliefs follows from KF-type arguments.
	
	For each player $i \in \cN$,  we define an unobserved state vector as
	\begin{subequations}
		\begin{equation}
			s^i_t=\left[\begin{array}{cc} v \ ; & \hv^{-i}_{t-1} \end{array}\right].
		\end{equation}
		and an observation vector
		\begin{equation}
			y^i_t=\left[\begin{array}{cc} a^{-i}_{t-1}-m^{-i}_{t-1} \ ; & x^i_t \end{array}\right].
		\end{equation}
	\end{subequations}
	
	We will show that the random vector $s^i_t$ evolves according to a linear Gaussian process,
	\begin{subequations}\label{lqgupdate}
		\begin{align}
			s^i_{t+1} &= \mat{A}^i_t s^i_t+\left[\begin{array}{c} \mat{0}\\ \mat{D}^i_t \end{array}\right]\vc{a}^i_{t-1}+\left[\begin{array}{c} \mat{0}\\ \mat{H}^i_t \end{array}\right]
			w^{-i}_t+\left[\begin{array}{c} \vc{0}\\ \vc{d}^i_t \end{array}\right]  \label{lqgupdate-a}\\
			y^i_t &=\mat{C}^i_t s^{i}_t+\left[\begin{array}{c}\mat{0}\\\mat{I}\end{array}\right]w^i_t,
		\end{align}
		where
		\begin{align}
			\mat{A}^i_t&=\left[\begin{array}{c} \begin{array}{cc}
					\mat{I} & \mat{0}
				\end{array}\\   \mat{G}^{-i}_t
			\end{array}	\right],\\
			\mat{C}^i_t&=\left[\begin{array}{cc}   \mat{0} & \mathfrak{D}(\mat{L}^{-i}_{t-1})\\
				\mat{I} & \mat{0}\end{array}\right].
		\end{align}
	\end{subequations}
	Note that $(y^i_{1:t},a^i_{1:t-1})$ is a shifted version of $h^i_t$. We prove the validity of~\eqref{lqgupdate} and the claim of the theorem using induction. In particular, Lemma~\ref{lm:indbase} below is the induction basis and the subsequent Lemma~\ref{lm:indstep} is the induction step.
	This concludes the proof of the theorem.

	

	\begin{lemma}\label{lm:indbase}
		The following are true.
		
		\noindent
		(a) $\xi^i_1$ is Gaussian $N(\hv^i_1,\mat{\Sigma}^i_1)$, with $\hv^i_1=\mat{\Sigma}{(\mat{\Sigma}+\mat{Q}^i)}^{-1}x^i_1$ and $\mat{\Sigma}^i_1=\mat{\Sigma}- \mat{\Sigma}(\mat{\Sigma}+\mat{Q}^i)^{-1}  \mat{\Sigma}$.
		Consequently the public belief $\pi^i_1(\xi^i_1|v)$ reduces to $\pi^i_1(\hv^i_1|v)$.
		
		\noindent
		(b)  \eqref{lqgupdate} holds for $t=1$.
		
		\noindent
		(c) The public belief $\pi^i_1(\hv^i_1|v)$ is Gaussian with mean
		$\E[\chv^i_1|v]=\mat{E}^{i}_1 v+\vc{f}^{i}_1$, with $\mat{E}^{i}_1=\mat{\Sigma}(\mat{\Sigma}+\mat{Q}^i)^{-1}$, $\vc{f}^{i}_1=0$, and covariance matrix $\mat{\Sigma}(\mat{\Sigma}+\mat{Q}^i)^{-1}\mat{Q}^i(\mat{\Sigma}+\mat{Q}^i)^{-1}\mat{\Sigma}$.
	\end{lemma}

	\begin{IEEEproof}
		
		\noindent
		(a) We have $x^i_1=v+w^i_1$ and $\xi^i_1(v)=\P(v|x^i_1)$, so due to joint Gaussianity of $V$ and $X^i_1$ we have that $\xi^i_1$
		is $N(\hv^i_1,\mat{\Sigma}^i_1)$, with mean
		\begin{subequations}
			\begin{align}
				\hv^i_1&=\E[V|x^i_1]\\&=\E[V]+\E[VX^{i\prime}_1]{\E[X^i_1X^{i\prime}_1]}^{-1}(x^i_1-\E[X^i_1])\\ &=\mat{\Sigma}{(\mat{\Sigma}+\mat{Q}^i)}^{-1}x^i_1,
				\label{basisV}
			\end{align}
		\end{subequations}
		and covariance matrix
		\begin{align}
			\mat{\Sigma}^i_1&=\mat{\Sigma}- \mat{\Sigma}(\mat{\Sigma}+\mat{Q}^i)^{-1}  \mat{\Sigma}.
			\label{basisCov}
		\end{align}
		As a result the only private information of user $i$ relevant to other users is $\hv^i_1$ and the public belief $\pi^i_1(\xi^i_1|v)$
		can be reduced to $\pi^i_1(\hv^i_1|v)$.

		\noindent
		(b) We have $s^i_1=\left[\begin{array}{cccc} v \ ; & \vc{0}\end{array}\right]$ and $s^i_2=\left[\begin{array}{cccc} v \ ; &  \hv^{-i}_1 \end{array}\right]$.
		The first row of~\eqref{lqgupdate-a} is evidently true.
		For the second row, using the result (from part (a)) $\hv^j_1=\mat{\Sigma}{(\mat{\Sigma}+\mat{Q}^j)}^{-1}(v+w^j_1)$, we can derive
		$\mat{G}^{-i}_1 $, $\mat{H}^i_1$, $\mat{D}^i_1$  and $\vc{d}^i_1$ as
		\begin{subequations}
			\begin{align}
				\mat{G}^{-i}_1 &=\left[\begin{array}{cc}  \mat{\Sigma}{(\mat{\Sigma}+\mat{Q}^{-i})}^{-1} & \mat{0} \end{array}\right]\\
				\mat{H}^i_1 &= \mathfrak{D}(\mat{\Sigma}{(\mat{\Sigma}+\mat{Q}^{-i})}^{-1})\\
				\mat{D}^i_1 &=\mat{0}\\
				\vc{d}^i_1 &=\mat{0},
			\end{align}
		\end{subequations}
		where $\mat{\Sigma}{(\mat{\Sigma}+\mat{Q}^{-i})}^{-1}$ is the vertical concatenation of the matrices $\mat{\Sigma}{(\mat{\Sigma}+\mat{Q}^j)}^{-1}$ for $j \in -i$.

		\noindent
		(c) Since $\hv^i_1=\mat{\Sigma}{(\mat{\Sigma}+\mat{Q}^i)}^{-1}(v+w^i_1)$ we deduce that $\pi^i_1(\hv^i_1|v)$ is Gaussian with mean $\E[\chv^i_1|v]=\mat{\Sigma}{(\mat{\Sigma}+\mat{Q}^i)}^{-1} v$ and covariance matrix $\tilde{\mat{\Sigma}}^i_1=\mat{\Sigma}(\mat{\Sigma}+\mat{Q}^i)^{-1}\mat{Q}^i(\mat{\Sigma}+\mat{Q}^i)^{-1}\mat{\Sigma}$.
		
	\end{IEEEproof}


	
	\begin{lemma}\label{lm:indstep}
		Assuming pure linear strategies of the form	$\gamma^j_t(a^j_t|\hv^j_t)=\delta(a^j_t-\mat{L}^j_t\hv^j_t-\vc{m}^j_t)$ for all $j \in \cN$, and assuming that~\eqref{lqgupdate} holds for $t\leq k$ and $\E[\chv^j_{k}|v,a_{1:k-1}]=\mat{E}^j_{k} v+\vc{f}^j_{k}$, the following are true.
		
		\noindent
		(a) $\xi^i_{k+1}$ is $N(\hv^i_{k+1},\mat{\Sigma}^i_{k+1})$ with
		\begin{align}
			\hv^i_{k+1}
			&= \mat{G}^{i,i}_{k+1}
			\left[\begin{array}{c} \hv^i_k\\ x^i_{k+1}
			\end{array}\right]
			+\vc{d}^{i,i}_{k+1},
		\end{align}
		where $\mat{G}^{i,i}_{k+1}$, $\vc{d}^{i,i}_{k+1}$ and $\mat{\Sigma}^i_{k+1}$ can be publicly evaluated.
		Consequently, the public belief $\pi^i_{k+1}(\xi^i_{k+1}|v)$ can be reduced to a belief $\pi^i_{k+1}(\hv^i_{k+1}|v)$.
		
		\noindent
		(b) \eqref{lqgupdate}  holds for $t=k+1$.
		
		\noindent
		(c) The conditional public belief, $\pi^i_{k+1}(\hv^i_{k+1}|v)$, are Gaussian with mean $\E[\chv^i_{k+1}|V,a_{1:k}]=\mat{E}^i_{k+1} V+\vc{f}^i_{k+1}$ and covariance matrix $\tilde{\mat{\Sigma}}^i_{k+1}$, where matrices $\mat{E}^i_{k+1}$ and $\tilde{\mat{\Sigma}}^i_{k+1}$ and vector $\vc{f}^i_{k+1}$
		can be publicly evaluated.
	\end{lemma}
	\begin{IEEEproof}
		\noindent
		(a) We first show one important result from the lemma assumptions. Notice that due to conditional independence of $x^j_{k}$'s given $v$ across time and players, and since $\hv^j_{k}$ is a function of $x^j_{1:k}$ and  $a_{1:k-1}$, we have
		\begin{subequations}
			\begin{align}
				\hhv^{i,j}_{k}&=\E[\chv^j_{k}|x^i_{1:k},a_{1:k-1}]\\
				&=\E_V[\E[\chv^j_{k}|V,x^i_{1:k},a_{1:k-1}]|x^i_{1:k},a_{1:k-1}]\\
				&=\E_V[\E[\chv^j_{k}|V,a_{1:k-1}]|x^i_{1:k},a_{1:k-1}]\\
				&= \E_V[\mat{E}^j_{k} V+\vc{f}^j_{k}|x^i_{1:k},a_{1:k-1}]\\
				&=\mat{E}^j_{k} \E[V|x^i_{1:k},a_{1:k-1}]+\vc{f}^j_{k}\\
				&=\mat{E}^j_{k}\hv^i_{k}+\vc{f}^j_{k}.
			\end{align}
			\label{cond}
		\end{subequations}
		By using the assumption that~\eqref{lqgupdate} holds for $t=k$,  we form a linear Gaussian model with partial observations and use Kalman filter results~\cite[Ch.7]{KuVa86}. Consider equation  \eqref{lqgupdate}  for $t=k$.
		%
		By using standard Kalman filter results~\cite[Ch.7]{KuVa86},
		we know that the belief over the system states given the observations is Gaussian and therefore, the private belief $\xi^i_k$ is $N(\hv^i_k,\mat{\Sigma}^i_k)$.
		We denote $\E[S^i_{k+1}|y^i_{1:k+1},a^i_{1:k}]$ and $\mathbb{E}
		[S^i_{k+1}|y^i_{1:k},a^i_{1:k-1}]$ by $s^i_{k+1|k+1}$ and $s^i_{k+1|k}$, respectively. We have
		\begin{subequations}
			\begin{align}	
				s^i_{k+1|k+1} &=\E[{S^i_{k+1}|x^i_{1:k+1},a_{1:k}}] \\
				&=\left[\hspace{-0.2cm}\begin{array}{c} \hv^i_{k+1} \\  \E[{\chv^{-i}_{k}|x^i_{1:k+1},a_{1:k}}] \end{array}\hspace{-0.2cm}\right]\\
				&= \mat{A}^i_{k}s^i_{k|k}+\left[\begin{array}{c} \mat{0}\\ \mat{D}^i_k \end{array}\right]\vc{a}^i_{k-1}\nonumber \\
				&\quad \ +\mat{J}^i_{k+1}( y^i_{k+1}-\mat{C}^i_{k+1}s^i_{k+1|k})+\left[\begin{array}{c} \vc{0}\\ \vc{d}^i_k \end{array}\right].
			\end{align}
		\end{subequations}
		Therefore,
		\begin{subequations}
			\begin{align}
				\hv^i_{k+1} &=\hv^i_{k}+(\mat{J}^i_{k+1})_{\hv^i,:}(y^i_{k+1}-\mat{C}^i_{k+1}s^i_{k+1|k})\\
				&= \hv^i_{k}+(\mat{J}^i_{k+1})_{\hv^i,:}\left[\begin{array}{c} a^{-i}_{k}-m^{-i}_{k}-\mathfrak{D}(\mat{L}^{-i}_{k})\hhv^{i,-i}_{k} \\x^i_{k+1} -\hv^i_k \end{array}\hspace{-0.1cm}\right]\hspace{-0.1cm}
				\\  &=\hv^i_{k}+(\mat{J}^i_{k+1})_{\hv^i,:}\left[\begin{array}{c} -\mathfrak{D}(\mat{L}^{-i}_{k})\mat{E}^{-i}_{k} \hv^i_{k} \\ x^i_{k+1} -\hv^i_{k} \end{array}\right]\nonumber \\
				&\quad + (\mat{J}^i_{k+1})_{\hv^i,a^{-i}} (a^{-i}_{k}-m^{-i}_{k}-\mathfrak{D}(\mat{L}^{-i}_{k})\vc{f}^{-i}_{k})  \\
				&= \mat{G}^{i,i}_{k+1} \left[\begin{array}{c} \hv^i_k\\ x^i_{k+1}  \end{array}\right]+\vc{d}^{i,i}_{k+1},
			\end{align}
			\label{vhupdate}
		\end{subequations}
		where
		\begin{subequations}
			\begin{align}
				(\mat{G}^{i,i}_{k+1})_{:,x^i} &=(\mat{J}^i_{k+1})_{\hv^i,x^i}\\
				(\mat{G}^{i,i}_{k+1})_{:,\hv^i} &= \mat{I}-\hspace{-0.05cm}(\mat{J}^i_{k+1})_{\hv^i,a^{-i}}\mathfrak{D}(\mat{L}^{-i}_{k})\mat{E}^{-i}_{k}  -(\mat{J}^i_{k+1})_{\hv^i,x^i} \\
				\vc{d}^{i,i}_{k+1} &= (\mat{J}^i_{k+1})_{\hv^i,a^{-i}} (a^{-i}_{k}-m^{-i}_{k}-\mathfrak{D}(\mat{L}^{-i}_{k})\vc{f}^{-i}_{k}).
			\end{align}
		\end{subequations}
The matrix  $\mat{J}^i_{k+1}$ and the covariance matrix of $\vc{S}^i_{k+1}$ conditioned on $y^i_{1:k+1}$ and $y^i_{1:k}$, denoted by $\mat{\Sigma}^i_{k+1|k+1}$ and $\mat{\Sigma}^i_{k+1|k}$, respectively, can be derived from the standard Kalman filter equations as follows
		\begin{subequations}
			\begin{align}
		\mat{\Sigma}^i_{k+1|k}
				&=\mat{A}^i_{k}\mat{\Sigma}^i_{k|k}\mat{A}^{i\prime}_{k}+\left[\begin{array}{c} \mat{0}\\ \mat{H}^i_k \end{array}\right] \mathfrak{D}(\mat{Q}^{-i}) \left[\begin{array}{c} \mat{0}\\ \mat{H}^i_k \end{array}\right]' 	\label{condsigma} \\
		\mat{J}^i_{k+1}
				&=\mat{\Sigma}^i_{k+1|k}\mat{C}^{i\prime}_{k+1}(\mat{C}^i_{k+1} \mat{\Sigma}^i_{k+1|k} \mat{C}^{i\prime}_{k+1}\nonumber\\
				& \qquad\qquad  +\left[\begin{array}{c}\mat{0} \\ \mat{I} \end{array}\right] \mat{Q}^i \left[\begin{array}{c}\mat{0} \\ \mat{I} \end{array}\right]' )^{-1}
				\label{J}\\
		\mat{\Sigma}^i_{k+1|k+1}
                &=(\mat{I}-\mat{J}^i_{k+1}\mat{C}^i_{k+1})\mat{\Sigma}^i_{k+1|k} 	\label{Sigma} \\
		\mat{\Sigma}^i_{1|1}
                &=\E[S^i_1S^{i\prime}_1]-\E[S^i_1 X^{i\prime}_1](\mathbb{E}
				[X^i_1 X^{i\prime}_1])^{-1}\E[S^i_1 X^{i\prime}_1]'\\
				&=\left[
				\begin{array}{cc} \mat{\Sigma} &  \mat{0} \\
					\mat{0}  &  \mat{0}
				\end{array}\right]-
				\left[\hspace{-0.1cm}\begin{array}{c} \mat{\Sigma} \\
					\mat{0}
				\end{array}\hspace{-0.1cm}\right]
				\hspace{-0.05cm}(\mat{\Sigma}+\mat{Q}^i)^{-1}\hspace{-0.05cm}\left[\hspace{-0.1cm}\begin{array}{cccc} \mat{\Sigma} & \mat{0}
				\end{array}\hspace{-0.1cm}\right]\\
				&=\left[
				\begin{array}{cccc}\mat{\Sigma}- \mat{\Sigma}(\mat{\Sigma}+\mat{Q}^i)^{-1}  \mat{\Sigma}& \mat{0} \\
					\mat{0} & \mat{0}
				\end{array}\right].
			\end{align}
			\label{sigma}
		\end{subequations}
		Note that, for notational simplicity, we remove the time subscripts from submatrix notation, so that $(\mat{J}^i_{k+1})_{\hv^i,x^i}$ denotes $(\mat{J}^i_{k+1})_{\hv^i_{k+1},x^i_k}$.
		
		Finally,  we have $\mat{\Sigma}^i_t=(\mat{\Sigma}^i_{t+1|t})_{v,v}$.
		Unlike $\hv^i_t$, which is part of  the private information of player $i$, the matrix $\mat{\Sigma}^i_t$ is a public quantity due to the independence of equation \eqref{Sigma} to the private observations of player $i$.

		
		\noindent
		(b)		Equation \eqref{lqgupdate-a} is obvious for the first part of the state, $v$. In order to prove the other parts of equation \eqref{lqgupdate-a} for $t=k+1$, we consider  the dynamic system \eqref{lqgupdate} for each of the  players $-i$ for $t=k$ and we write \eqref{vhupdate} for players $-i$. Since $x^{-i}_{k+1}$ is not part of $y^i_{k+1}$, we can substitute it by $v+w^{-i}_{k+1}$ and derive $\mat{G}^{j}_{k+1}$, $\mat{D}^{i}_{k+1}$, $\mat{H}^i_{k+1}$, and $\vc{d}^i_{k+1}$ for all $j \in -i$ as
		\begin{subequations}
			\begin{align}
	(\mat{G}^{j}_{k+1})_{:,v}
                &=(\mat{J}^j_{k+1})_{\hv^j,x^j} \\
	(\mat{G}^{j}_{k+1})_{:,\hv^j}
				&=\mat{I}-(\mat{J}^j_{k+1})_{\hv^j,a^{-j}}\mathfrak{D}(\mat{L}^{-j}_{k})\mat{E}^{-j}_{k}
				-(\mat{J}^j_{k+1})_{\hv^j,x^j} \\
	(\mat{D}^i_{k+1})_{\hv^j,:}
                &=(\mat{J}^j_{k+1})_{\hv^j,a^{i}} \\
	(\vc{d}^i_{k+1})_{\hv^j}
				&= (\mat{J}^j_{k+1})_{\hv^j,a^{-ij}}
                    (a^{-ij}_{k}-m^{-ij}_{k}-\mathfrak{D}(\mat{L}^{-ij}_{k})\vc{f}^{-ij}_{k})\nonumber \\
				&\qquad  +(\mat{J}^j_{k+1})_{\hv^j,a^{i}} (-m^i_{k}-\mat{L}^{i}_{k}\vc{f}^{i}_{k})\\
	\mat{H}^{i}_{k+1}
				&= \mathfrak{D}((\mat{J}^{-i}_{k+1})_{\hv^{-i},x^{-i}}).
			\end{align}
		\end{subequations}
		%
		%
		The notation $-ij$ means all of the players exept $i$ and $j$. We have derived the matrices $\mat{A}^i_{k+1}$, $\mat{D}^i_{k+1}$, $\mat{H}^i_{k+1}$,  and vector $\vc{d}^i_{k+1}$ and so \eqref{lqgupdate} holds for $t=k+1$.
		
		
		\noindent
		(c)	In order to show that the conditional public belief $\pi^i_{k+1}(\hv^i_{k+1}|v)$ is Gaussian, we consider a conditional  Gauss Markov model.
		Note that the conditional public belief  is publicly measurable conditioned on $V$. We use this fact to form a conditional model, where the observations are the conditions in the conditional public belief and we derive conditional Kalman filters.
		Using \eqref{lqgupdate} for $t\leq k+1$, we can construct the following linear Gaussian model for $t\leq k+1$,
		\begin{subequations}
			\begin{align}
				& \hspace{-1.6cm}\text{State:} \nonumber \\
				&\tilde{s}_{t}=\left[\begin{array}{c} v\\ \hv_{t-1} \end{array}\right], \\
				&\hspace{-1.6cm}\text{State Evolution:} \nonumber \\
				&\tilde{s}_{t+1} = \tilde{\mat{A}}_t \tilde{s}_t +\tilde{\mat{H}}_t w_t+\tilde{\vc{d}}_t,\\
				&\hspace{-1.6cm}\text{Observation:} \nonumber \\
				&\tilde{y}_t=\left[\begin{array}{c}v \\ a_{t-1}-m_{t-1} \end{array}\right]=\tilde{\mat{C}}_ts_t,
			\end{align}
		\end{subequations}
		where
		\begin{subequations}
			\begin{align}
				\tilde{\mat{A}}_t &=\left[\begin{array}{cc} \mat{I}  & \mat{0} \\ & \hspace{-0.5cm}  \tilde{\mat{G}}_t \end{array}\right]\\
				(\tilde{\mat{G}}_t)_{\hv^i,v\hv^i} &=(\mat{G}^i_t)_{:,v\hv^i}, \quad \forall i \in \cN\\
				(\tilde{\mat{H}}_t)_{\hv^i,w^i} &=(\mat{J}^i_{t})_{\hv^i,x^i}, \quad \forall i \in \cN\\
				(\tilde{\vc{d}}_t)_{\hv^i} &=d^{i,i}_t, \quad \forall i \in \cN\\
				\tilde{\mat{C}}_t &=\left[\begin{array}{cc}\mat{I} &  \mat{0} \\ \mat{0} & \quad \mathfrak{D}(\mat{L}_{t-1})  \end{array}\right].
			\end{align}
		\end{subequations}
		Using this conditional Gauss Markov model, we can conclude that the conditional public beliefs $\pi^j_{k+1}(\hv^j_{k+1}|v)$ are Gaussian and  by using Kalman filter results for $t=k+1$, we can write
		\begin{subequations}
			\begin{align}
				\tilde{s}&_{k+2|k+1} \nonumber \\
				&=\E[\tilde{S}_{k+2}|\tilde{y}_{1:k+1}] \\
				&=\E[\tilde{S}_{k+2}|v,a_{1:k}]  \\ &=\tilde{\mat{A}}_{k+1}\tilde{s}_{k+1|k}+\tilde{\mat{A}}_{k+1}\tilde{\mat{J}}_{k+1}(\tilde{y}_{k+1}-\tilde{\mat{C}}_{k+1}\tilde{s}_{k+1|k})+\tilde{\vc{d}}_{k+1}.
			\end{align}
		\end{subequations}
		Therefore,
		\begin{subequations}
			\begin{align}
				\E[&\chv_{k+1}|v,a_{1:k}]\nonumber \\&=(\tilde{\mat{G}}_{k+1})_{:,v}v +(\tilde{\mat{G}}_{k+1})_{:,\hv} \E[\chv_{k}|v,a_{1:k-1}]  \nonumber   \\&\quad -(\tilde{\mat{A}}_{k+1}\tilde{\mat{J}}_{k+1})_{\hv,a}\mathfrak{D}(\mat{L}_{k})\E[\chv_{k}|v,a_{1:k-1}]\nonumber  \\& \quad
				+(\tilde{\mat{A}}_{k+1}\tilde{\mat{J}}_{k+1})_{\hv,a}(a_{k}-m_{k})+(\tilde{\vc{d}}_{k+1})_{\hv}.
			\end{align}
		\end{subequations}
		Using the assumption of $\E[\chv_{k}|v,a_{1:k-1}]= \mat{E}_{k}v+\vc{f}_{k}$, we have the following
		\begin{subequations}
			\begin{align} \E[&\chv_{k+1}|v,a_{1:k}] \nonumber \\
				&=(\tilde{\mat{G}}_{k+1})_{:,v}v+(\tilde{\mat{G}}_{k+1})_{:,\hv}(\mat{E}_{k}v+\vc{f}_{k}) \nonumber  \\&\quad -(\tilde{\mat{A}}_{k+1}\tilde{\mat{J}}_{k+1})_{\hv,a}\mathfrak{D}(\mat{L}_{k})(\mat{E}_{k}v+\vc{f}_{k})\nonumber  \\&
				\quad +(\tilde{\mat{A}}_{k+1}\tilde{\mat{J}}_{k+1})_{\hv,a}(a_{k}-m_{k})+(\tilde{\vc{d}}_{k+1})_{\hv}\\
				& = \mat{E}_{k+1} v+\vc{f}_{k+1},
			\end{align}
			
		\end{subequations}
		
		where
		\begin{subequations}
			\begin{align}
				\mat{E}_{k+1} &=(\tilde{\mat{G}}_{k+1})_{:,v}+((\tilde{\mat{G}}_{k+1})_{:,\hv}\nonumber \\
								&\quad -(\tilde{\mat{A}}_{k+1}\tilde{\mat{J}}_{k+1})_{\hv,a}\mathfrak{D}(\mat{L}_{k}))\mat{E}_{k}\\
				\vc{f}_{k+1}&=((\tilde{\mat{G}}_{k+1})_{:,\hv}-(\tilde{\mat{A}}_{k+1}\tilde{\mat{J}}_{k+1})_{\hv,a}\mathfrak{D}(\mat{L}_{k}))\vc{f}_{k}\nonumber \\&\quad +(\tilde{\mat{A}}_{k+1}\tilde{\mat{J}}_{k+1})_{\hv,a}(a_{k}-m_{k})+(\tilde{\vc{d}}_{k+1})_{\hv},
				\label{EFupdate}
			\end{align}
		\end{subequations}
		and similar to part (a) of the proof, the covariance matrix of $\tilde{\vc{S}}_{k+1}$ conditioned on $\tilde{y}_{1:k+1}$ and $\tilde{y}_{1:k}$, denoted by $\tilde{\mat{\Sigma}}_{k+1|k+1}$ and   $\tilde{\mat{\Sigma}}_{k+1|k}$, respectively,  and the matrix $\tilde{\mat{J}}_{k+1} $ are derived from the following Kalman filter equations.
		\begin{subequations}
			\begin{align}  \tilde{\mat{\Sigma}}_{k+1|k} &=\tilde{\mat{A}}_{k}\tilde{\mat{\Sigma}}_{k|k}\tilde{\mat{A}}^{'}_{k}+\tilde{\mat{H}}_{k} \mathfrak{D}(\mat{Q})\tilde{ \mat{H}}^{'}_{k} \label{Jtilde} \\
				\tilde{\mat{J}}_{k+1} &=\tilde{\mat{\Sigma}}_{k+1|k}\tilde{\mat{C}}^{'}_{k+1}(\tilde{\mat{C}}_{k+1} \tilde{\mat{\Sigma}}_{k+1|k} \tilde{\mat{C}}^{'}_{k+1} )^{-1}
				\label{condsigmatilde}\\
				\tilde{\mat{\Sigma}}_{k+1|k+1} &=(\mat{I}-\tilde{\mat{J}}_{k+1}\tilde{\mat{C}}_{k+1})\tilde{\mat{\Sigma}}_{k+1|k} \label{Sigmatilde}\\
				\tilde{\mat{\Sigma}}_{1|1} &=\E[\tilde{S}_1\tilde{S}^{'}_1]-\E[\tilde{S}_1 V'](\E[VV'])^{-1}\E[\tilde{S}_1 V']' \\
				&=\mat{0}.
			\end{align}
			\label{sigmatilde}
		\end{subequations}
		
		Note that if we know $\mat{\Sigma}_{k+1|k}$, $\tilde{\mat{\Sigma}}_{k+1|k}$, $\mat{E}_k$  and $\vc{f}_k$, we can publicly evaluate all of the other quantities defined in this proof for $k+1$ for a given strategy matrices $\mat{L}_k$ and vectors $\vc{m}_k$  and therefore, we can find $\mat{\Sigma}_{k+2|k+1}$, $\tilde{\mat{\Sigma}}_{k+2|k+1}$, $\mat{E}_{k+1}$ and $\vc{f}_{k+1}$. We can also find $\mat{G}^{i,i}_{k+1}$ and $d^{i,i}_{k+1}$, which are used to update $\hv^i_k$ to $\hv^i_{k+1}$.
			\end{IEEEproof}

		\section{Proof of Theorem \ref{thm:linstlqg}}\label{prf:linstlqg}
	We show that for any $t \in \cT$, if all players $-i$ play according to the strategy $\gamma^{-i}_t(a^{-i}_t|\hv^{-i}_t)=\delta(a^{-i}_t-\mathfrak{D}(\mat{L}^{-i}_t)\hv^{-i}_t-m^{-i}_t)$, where $m^{-i}_t=\mat{M}^{-i}_tf_t+\bar{m}^{-i}_t$, and the strategies of players are linear in $\hv_k$ for $k < t$, player $i$ faces an MDP with state $(\hv^i_t,\mat{\Sigma}_t,\mat{E}_t,\vc{f}_t)$ and her best response is of the form $\gamma^{i}_t(a^{i}_t|\hv^{i}_t)=\delta(a^{i}_t-\mathfrak{D}(\mat{L}^{i}_t)\hv^{i}_t-m^{i}_t)$, where $m^{i}_t=\mat{M}^{i}_tf_t+\bar{m}^{i}_t$.
	
	By using the results from Theorem \ref{thm:lqg-belief}, given the strategy profile $\gamma_t$, $(\hv^i_t,\mat{\Sigma}_t,\mat{E}_t,\vc{f}_t)$ forms a Markov chain. Notice that $\chv^i_{t+1},\mat{\Sigma}_{t+1},\mat{E}_{t+1},\vc{f}_{t+1}$ are updated by $\gamma_t$ which is linear and therefore, all  results from Theorem \ref{thm:lqg-belief} hold.
	
	\begin{lemma}\label{lm:lastlemma}
One can write the expected value of the instantaneous reward $\bar{R}^i_t=\E[r^i_t(V,A_t)|a^i_t,\hv^i_t,\mat{\Sigma}^i_t,\mat{E}_t,\vc{f}_t]$  as
		\begin{align}
		\bar{R}^i_t=\qd(\bar{\mat{R}}^i_t;\left[\begin{array}{c}
		\hv^i_t \\ a^i_t \\ \vc{f}_t
		\end{array}\right])
		+{\bar{\vc{b}}^{i\prime}_t}\left[\begin{array}{c}
		\hv^i_t \\ a^i_t \\ f_t
		\end{array}\right]+\bar{\vc{c}}^i_t,
		\label{rewquad}
		\end{align}

		where $\bar{\mat{R}}^i_t$, $\bar{b}^i_t$ and $\bar{c}^i_t$ are constructed in the proof.
	\end{lemma}
	\begin{IEEEproof}
		Since we assume all players $-i$ play according to $\gamma^{-i}_t$, we have $a^{-i}_t=\mathfrak{D}(\mat{L}^{-i}_t)\hv^{-i}_t+\mat{M}^{-i}_tf_t+\bar{m}^{-i}_t$ and so  the instantaneous reward can be rewritten as follows.
		\begin{subequations}
			\begin{align}
			r^i_t(v,a_t)&=\qd(\mat{R}^i_t; \left[\begin{array}{c}
			v \\ a_t
			\end{array}\right]) \\&=\qd (\tilde{\mat{R}}^i_t;\left[\begin{array}{c}
			v \\ a^i_t \\ \hv^{-i}_t \\f_t
			\end{array}\right])+\tilde{\vc{b}}^{i\prime}_t\left[\begin{array}{c}
			v \\ a^i_t \\ \hv^{-i}_t \\ f_t
			\end{array}\right]+\tilde{c}^i_t,
			\end{align}
		\end{subequations}
		
		where
		\begin{subequations}
			\begin{align}
			&\tilde{\mat{R}}^i_t=\left[\begin{array}{ccc}
			\mat{I}_{N_v+N_a} & \mat{0} & \mat{0}\\ \mat{0} & \mathfrak{D}(\mat{L}^{-i}_t) & \mat{M}^{-i}_t
			\end{array}\right]'\tilde{\mat{I}}_{2,i+1}' \mat{R}^i_t\tilde{\mat{I}}_{2,i+1}\\&\hspace{4cm}\left[\begin{array}{ccc}
			\mat{I}_{N_v+N_a} & \mat{0} &  \mat{0}\\ \mat{0} & \mathfrak{D}(\mat{L}^{-i}_t) & \mat{M}^{-i}_t
			\end{array}\right]\\
			&\tilde{\mat{I}}_{2,i+1}=\left[\begin{array}{cccc}
			\mat{I}_{N_v} & \mat{0} &  \mat{0}  &  \mat{0}  \\ \mat{0} &  \mat{0}& \mat{I}_{(i-1)N_a} &  \mat{0}\\
			\mat{0} & \mat{I}_{N_a} & \mat{0} &\mat{0}\\
			\mat{0} & \mat{0} & \mat{0} &  \mat{I}_{(N-i)N_a}
			\end{array}\right],
			\end{align}
			where $\mat{I}_k$ is the identity matrix with size $k\times k$.
			\begin{align}
			& \tilde{\vc{b}}^{i\prime}_t=2\left[\begin{array}{c}
			\vc{0} \\ \bar{\vc{m}}^{-i}_t
			\end{array}\right]'\tilde{\mat{I}}_{2,i+1}' \mat{R}^i_t\tilde{\mat{I}}_{2,i+1}\nonumber \\
			& \hspace{3cm}\left[\begin{array}{ccc}
			\mat{I}_{N_v+N_a} & \mat{0} & \mat{0}\\ \mat{0} & \mathfrak{D}(\mat{L}^{-i}_t) & \mat{M}^{-i}_t
			\end{array}\right]\\
			&\tilde{\vc{c}}^i_t=\left[\begin{array}{c}
			\vc{0} \\ \bar{\vc{m}}^{-i}_t
			\end{array}\right]'\tilde{\mat{I}}_{2,i+1}' \mat{R}^i_t\tilde{\mat{I}}_{2,i+1}\left[\begin{array}{c}
			\vc{0} \\ \bar{\vc{m}}^{-i}_t
			\end{array}\right].
			\end{align}
		\end{subequations}
		We can now calculate the expected value of $R^i$ as follows.
		\begin{align}
		\bar{R}^i_t=\qd(\tilde{\mat{R}}^i_t; \left[\begin{array}{c}
		\hv^i_t \\ a^i_t\\\hhv^{i,-i}_t\\f_t
		\end{array}\right])+\tr(\tilde{\mat{R}}^i_t\bar{\mat{\Sigma}}^i_t)+\tilde{\vc{b}}^{i\prime}_t\left[\begin{array}{c}
		\hv^i_t \\ a^i_t \\ \hhv^{i,-i}_t\\f_t
		\end{array}\right]+\tilde{c}^i_t,
		\end{align}
		where
		\begin{align}
		\bar{\mat{\Sigma}}^i_t=\left[\begin{array}{cccc}
		\mat{\Sigma}^i_t & \mat{0} &\mat{\Sigma}^i_t \mat{E}^{-i'}_t  & \mat{0}\\ \mat{0} &  \mat{0}&  \mat{0} & \mat{0}\\
		\mat{E}^{-i}_t  \mat{\Sigma}^{i}_t & \mat{0} &  (\mat{\Sigma}^i_{t+1|t})_{\hv^{-i},\hv^{-i}} & \mat{0}\\
		\mat{0} &  \mat{0}&  \mat{0} & \mat{0}
		\end{array}\right].
		\end{align}
		By using $\hhv^{i,-i}_t=\mat{E}^{-i}_t \hv^i_t+\vc{f}^{-i}_t$, we can derive the equations for $\bar{\mat{R}}^i_t$, $\bar{\vc{b}}^i_t$ and $\bar{\vc{c}}^i_t$.
		\begin{subequations}
			\begin{align}
			&	\bar{\mat{R}}^i_t =
			\left[\begin{array}{ccc}
			\mat{I}_{N_v}&  \mat{0} & \mat{0} \\ \mat{0} & 	\mat{I}_{N_a} & \mat{0} \\
			\mat{E}^{-i}_t & \mat{0} & \hat{\mat{I}}_{-i}\\ \mat{0} & \mat{0} & \mat{I}_{N_vN}
			\end{array}\right]'\tilde{\mat{R}}^i_t 	\left[\begin{array}{ccc}
			\mat{I}_{N_v}&  \mat{0} & \mat{0} \\ \mat{0} & 	\mat{I}_{N_a} & \mat{0} \\
			\mat{E}^{-i}_t & \mat{0} & \hat{\mat{I}}_{-i}\\ \mat{0} & \mat{0} & \mat{I}_{N_vN}
			\end{array}\right]\\
			& \bar{b}^{i\prime}_t= \tilde{b}^{i\prime}_t	\left[\begin{array}{ccc}
			\mat{I}_{N_v}&  \mat{0} & \mat{0} \\ \mat{0} & 	\mat{I}_{N_a} & \mat{0} \\
			\mat{E}^{-i}_t & \mat{0} & \hat{\mat{I}}_{-i}\\ \mat{0} & \mat{0} & \mat{I}_{N_vN}
			\end{array}\right]\\
			& (\hat{\mat{I}}_{-i})_{:,f^{-i}}=\mat{I}_{(N-1)N_v}\\
			& \bar{c}^i_t=\tr(\tilde{\mat{R}}^i_t\bar{\mat{\Sigma}}^i_t)+\tilde{c}^i_t,
			\end{align}
		\end{subequations}
		
	\end{IEEEproof}
	In the next lemma, we show that the reward-to-go at time $t$ is a  quadratic functions of $\left[\begin{array}{c}
	\hv^i_t \\ f_t
	\end{array}\right]$ and we will construct the strategy matrix and vector $\mat{L}^i_t$ and $\vc{m}^i_t$.
	\begin{lemma}
		We have the following equation for the reward-to-go function, $J^i_t(\hv^i_t,\mat{\Sigma}_t,\mat{E}_t,\vc{f}_t)=\qd(\mat{Z}^i_t;\left[\begin{array}{c}
		\hv^i_t\\\vc{f}_t
		\end{array}\right])+\vc{z}^{i\prime}_t\left[\begin{array}{c}
		\hv^i_t\\\vc{f}_t
		\end{array}\right]+o^i_t.$
		\label{retogoquad}
	\end{lemma}
	Note that the above equation only highlights the functionality of the reward-to-go with respect to $\hv^i_t$ and $\vc{f}_t$. We do not care about its functionality with respect to $\mat{\Sigma}_t$ and $\mat{E}_t$ due to two reasons. First, they are part of the public part of the history and are not parameters of the partial strategies $\gamma$. Second, they are not controlled by the actions. As we will see in the proof of this lemma, $\mat{Z}^i_t$, $\vc{z}^i_t$ and $\vc{o}^i_t$ are  functions of $\mat{\Sigma}_t$ and $\mat{E}_t$.
	
	\begin{IEEEproof}
		We prove the lemma by backward induction. For $T+1$, we have $J^i_{T+1}(\hv^i_{T+1},\mat{\Sigma}_{T+1},\mat{E}_{T+1},\vc{f}_{T+1})=0$ and by setting $\mat{Z}^i_{T+1}=\mat{0}$, $\vc{z}^i_{T+1}=0$, $o^i_{T+1}=0$, the equation holds.
		
		Assume that the lemma holds for $t+1$. We will show that it will also hold for $t$.
		\begin{subequations}
			\begin{align}
			J^i_t&(\hv^i_t,\mat{\Sigma}_t,\mat{E}_t,\vc{f}_t)=\max_{a^i_t} \E^{\gamma^{-i}_t}[r^i_t(V,A_t) \nonumber \\
			&\quad +J^i_{t+1}(\chv^i_{t+1},\mat{\Sigma}_{t+1},\mat{E}_{t+1},\vc{f}_{t+1})|a^i_t,\hv^i_t,\mat{\Sigma}_t,\mat{E}_t,\vc{f}_t]  \\
			& =\max_{a^i_t} \{ \qd(\bar{\mat{R}}^i_t;\left[\begin{array}{c}
			\hv^i_t \\ a^i_t \\ \vc{f}_t
			\end{array}\right])
			+{\bar{\vc{b}}^{i\prime}_t}\left[\begin{array}{c}
			\hv^i_t \\ a^i_t \\ f_t
			\end{array}\right]+\bar{\vc{c}}^i_t  \nonumber \\&
			\quad +\E^{\gamma^{-i}_t}[ \qd(\mat{Z}^i_{t+1};\left[\begin{array}{c}
			\chv^i_{t+1}\\\vc{f}_{t+1}
			\end{array}\right])+\vc{z}^{i\prime}_{t+1}\left[\begin{array}{c}
			\chv^i_{t+1}\\\vc{f}_{t+1}
			\end{array}\right] \nonumber \\
			&\quad +o^i_{t+1}|a^i_t,\hv^i_t,\mat{\Sigma}_t,\mat{E}_t,\vc{f}_t]\}.
			\end{align}
		\end{subequations}
		
		First consider the $J^i_{t+1}$ part.
		\begin{align}
		&\E^{\gamma^{-i}_t}[ \qd(\mat{Z}^i_{t+1};\left[\begin{array}{c}
		\chv^i_{t+1}\\\vc{f}_{t+1}
		\end{array}\right])+\vc{z}^{i\prime}_{t+1}\left[\begin{array}{c}
		\chv^i_{t+1}\\\vc{f}_{t+1}
		\end{array}\right] \nonumber \\
		&\quad +o^i_{t+1}|a^i_t,\hv^i_t,\mat{\Sigma}_t,\mat{E}_t,\vc{f}_t] \nonumber \\&
		=\E^{\gamma^{-i}_t}[ \qd(\mat{Z}^i_{t+1};\hat{\mat{G}}^i_{t+1}\left[\begin{array}{c}
		\hv^i_{t}\\ a^i_t\\ \chv^{-i}_t \\ X^i_{t+1} \\ \vc{f}_t
		\end{array}\right]+\hat{\vc{g}}^i_{t+1}) \nonumber\\&
		\quad +\vc{z}^{i\prime}_{t+1}(\hat{\mat{G}}^i_{t+1}\left[\begin{array}{c}
		\hv^i_{t}\\ a^i_t\\ \chv^{-i}_t \\ X^i_{t+1} \\ \vc{f}_t
		\end{array}\right]+\hat{\vc{g}}^i_{t+1}) \nonumber \\
		&\quad +o^i_{t+1}|a^i_t,\hv^i_t,\mat{\Sigma}_t,\mat{E}_t,\vc{f}_t]\\
		&=\qd(\bar{\mat{Z}}^i_{t+1};\left[\begin{array}{c}
		\hv^i_t \\ a^i_t \\ \vc{f}_t
		\end{array}\right])+\bar{\vc{z}}^{i\prime}_{t+1}\left[\begin{array}{c}
		\hv^i_t \\ a^i_t \\ \vc{f}_t
		\end{array}\right]+\bar{o}^i_{t+1},
		\end{align}
		where
		\optv{submission}{the definition of $\hat{\mat{G}}^i_{t+1}$ is omitted due to space limitations and can be found in \cite{HeAn20arxiv}, and for the rest of the quantities we have}
			\optv{arxiv}{
		\begin{subequations}
			\begin{align}
			&(\hat{\mat{G}}^i_{t+1})_{\hv^i,\hv^i}=(\mat{G}^{i,i}_{t+1})_{:,\hv^i}\\
			& (\hat{\mat{G}}^i_{t+1})_{\hv^i,\hv^{-i}}= (\mat{J}^i_{t+1})_{\hv^i,a^{-i}} \mathfrak{D}(\mat{L}^{-i}_t) \\
			&(\hat{\mat{G}}^i_{t+1})_{\hv^i,x^i}=(\mat{G}^{i,i}_{t+1})_{:,x^i}\\
			& (\hat{\mat{G}}^i_{t+1})_{\hv^i,\vc{f}^{-i}}= (\mat{J}^i_{t+1})_{\hv^i,a^{-i}} \mathfrak{D}(\mat{L}^{-i}_t)\\
			& (\hat{\mat{G}}^i_{t+1})_{f^j,\vc{f}^{-j}} = ((\tilde{\mat{G}}_{t+1})_{:,\hv}-(\tilde{\mat{A}}_{t+1}\tilde{\mat{J}}_{t+1})_{\hv,a}\mathfrak{D}(\mat{L}_t))_{f^j,f^{-j}} \nonumber \\
			&\hspace{0.4cm}  -(\mat{J}^j_{t+1})_{\hv^j,a^{-j}}\mathfrak{D}(\mat{L}^{-j}_t) -(\tilde{\mat{A}}_{t+1}\tilde{\mat{J}}_{t+1})_{\hv^j,a^i}(\mat{M}^i_t)_{:,f^{-j}} \nonumber \\&\hspace{0.4cm}-(\mat{J}^j_{t+1})_{\hv^j,a^{i}}(\mat{M}^i_t)_{:,f^{-j}}, \ \forall j \neq i\\
			& (\hat{\mat{G}}^i_{t+1})_{f^j,\vc{f}^{j}}= ((\tilde{\mat{G}}_{t+1})_{:,\hv}-(\tilde{\mat{A}}_{t+1}\tilde{\mat{J}}_{t+1})_{\hv,a}\mathfrak{D}(\mat{L}_t))_{f^j,f^{j}} \nonumber \\&\qquad -(\tilde{\mat{A}}_{t+1}\tilde{\mat{J}}_{t+1})_{\hv^j,a^i}(\mat{M}^i_t)_{:,f^{j}} -(\mat{J}^j_{t+1})_{\hv^j,a^{i}}(\mat{M}^i_t)_{:,f^{j}}, \nonumber\\& \hspace{6.5cm} \forall j \neq i \\
			& (\hat{\mat{G}}^i_{t+1})_{f^i,\vc{f}^{-i}} = ((\tilde{\mat{G}}_{t+1})_{:,\hv}-(\tilde{\mat{A}}_{t+1}\tilde{\mat{J}}_{t+1})_{\hv,a}\mathfrak{D}(\mat{L}_t))_{f^i,f^{-i}} \nonumber \\
			&\qquad  -(\mat{J}^i_{t+1})_{\hv^i,a^{-i}}\mathfrak{D}(\mat{L}^{-i}_t)-(\tilde{\mat{A}}_{t+1}\tilde{\mat{J}}_{t+1})_{\hv^i,a^i}(\mat{M}^i_t)_{:,f^{-i}} , \\
			& (\hat{\mat{G}}^i_{t+1})_{f^i,\vc{f}^{i}}= ((\tilde{\mat{G}}_{t+1})_{:,\hv}-(\tilde{\mat{A}}_{t+1}\tilde{\mat{J}}_{t+1})_{\hv,a}\mathfrak{D}(\mat{L}_t))_{f^i,f^{i}} \nonumber \\&\qquad \qquad \qquad -(\tilde{\mat{A}}_{t+1}\tilde{\mat{J}}_{t+1})_{\hv^i,a^i}(\mat{M}^i_t)_{:,f^{i}},  \  \forall j \neq i \\
			& (\hat{\mat{G}}^i_{t+1})_{f^j,a^i}= (\mat{J}^j_{t+1})_{\hv^j,a^{i}}+(\tilde{\mat{A}}_{t+1}\tilde{\mat{J}}_{t+1})_{\hv^j,a^i}, \forall j \neq i\\
			& (\hat{\mat{G}}^i_{t+1})_{f^i,a^i}= (\tilde{\mat{A}}_{t+1}\tilde{\mat{J}}_{t+1})_{\hv^i,a^i}\\
			& (\hat{\mat{G}}^i_{t+1})_{f^k,\hv^j}= (\mat{J}^k_{t+1})_{\hv^k,a^{j}}\mat{L}^j_t+(\tilde{\mat{A}}_{t+1}\tilde{\mat{J}}_{t+1})_{\hv^k,a^j}\mat{L}^j_t,\nonumber \\& \hspace{5.4cm} \ \forall j \neq i, \forall k \neq j\\
			& (\hat{\mat{G}}^i_{t+1})_{f^j,\hv^j}=(\tilde{\mat{A}}_{t+1}\tilde{\mat{J}}_{t+1})_{\hv^j,a^j}\mat{L}^j_t, \  \forall j \neq i\\
			& (\hat{g}^i_{t+1})_{\vc{f}^i}=-(\tilde{\mat{A}}_{t+1}\tilde{\mat{J}}_{t+1})_{\hv^i,a^i}\bar{\vc{m}}^i_t\\
			& (\hat{g}^i_{t+1})_{\vc{f}^j}=-(\tilde{\mat{A}}_{t+1}\tilde{\mat{J}}_{t+1})_{\hv^j,a^i}\bar{\vc{m}}^i_t-(\mat{J}^j_{t+1})_{\hv^j,a^{i}}\bar{\vc{m}}^i_t,  \ \forall j \neq i,
			\end{align}
		\end{subequations}
	and we have}
		\begin{subequations}
			\begin{align}
			&\bar{\mat{Z}}^i_{t+1}=\mat{T}^{i\prime}_{t+1}\hat{\mat{G}}^{i\prime}_{t+1}\mat{Z}^i_{t+1}\hat{\mat{G}}^i_{t+1}\mat{T}^i_{t+1}\\
			&\mat{T}^i_{t+1}=\left[\begin{array}{ccc}
			\mat{I}_{N_v} & \mat{0} & \mat{0}\\
			\mat{0} & \mat{I}_{N_a} & \mat{0}\\
			\mat{E}^{-i}_t & \mat{0} & \hat{\mat{I}}_{-i}\\
			\mat{I}_{N_v}  & \mat{0} & \mat{0}\\
			\mat{0} & \mat{0} & \mat{I}_{N_vN}
			\end{array}\right]\\
			& \bar{\vc{z}}^{i\prime}_{t+1}=(2\hat{\vc{g}}^{i\prime}_{t+1}\mat{Z}^i_{t+1}\hat{\mat{G}}^i_{t+1}+\vc{z}^{i\prime}_{t+1}\hat{\mat{G}}^i_{t+1})\mat{T}^i_{t+1}\\
			& \bar{o}^i_{t+1}=\hat{\vc{g}}^{i\prime}_{t+1}\mat{Z}^i_{t+1}\hat{\vc{g}}^i_{t+1}+\tr(\hat{\mat{G}}^{i\prime}_{t+1}\mat{Z}^i_{t+1}\hat{\mat{G}}^i_{t+1}\hat{\mat{\Sigma}}^i_{t+1})\nonumber\\&\qquad \ \ +\vc{z}^{i\prime}_{t+1}\hat{\vc{g}}^i_{t+1}+o^i_{t+1}\\
			&\hat{\mat{\Sigma}}^i_{t+1}=Cov(\left[\begin{array}{c}
			\hv^i_{t}\\ a^i_t\\ \chv^{-i}_t \\ X^i_{t+1} \\ \vc{f}_t
			\end{array}\right]|a^i_t,\hv^i_t,\mat{\Sigma}_t,\mat{E}_t,\vc{f}_t)\\
			&(\hat{\mat{\Sigma}}^i_{t+1})_{\hv^{-i}x^i,\hv^{-i}x^i}=\left[\begin{array}{cc} (\mat{\Sigma}^i_{t+1|t})_{\hv^{-i},\hv^{-i}} & \mat{E}^{-i}_t \mat{\Sigma}^i_t\\
			\mat{\Sigma}^i_t\mat{E}^{-i'}_t  & \mat{\Sigma}^i_t+ \mat{Q}^i
			\end{array}\right]\hspace{-0.05cm}.
			\end{align}
		\end{subequations}
		
		Therefore, one can write the expected reward-to-go as follows.
		\begin{align}
		J^i_t&(\hv^i_t,\mat{\Sigma}^i_t,\mat{E}_t,\vc{f}_t)=\max_{a^i_t} \{ \qd(\bar{\mat{R}}^i_t;\left[\begin{array}{c}
		\hv^i_t \\ a^i_t \\ \vc{f}_t
		\end{array}\right])
		+\bar{\vc{b}}^{i\prime}_t\left[\begin{array}{c}
		\hv^i_t \\ a^i_t \\ f_t
		\end{array}\right]\nonumber  \\&  +\bar{\vc{c}}^i_t+
		\qd(\bar{\mat{Z}}^i_{t+1};\left[\begin{array}{c}
		\hv^i_t \\ a^i_t \\ \vc{f}_t
		\end{array}\right])+\bar{\vc{z}}^{i\prime}_{t+1}\left[\begin{array}{c}
		\hv^i_t \\ a^i_t \\ \vc{f}_t
		\end{array}\right]+\bar{o}^i_{t+1}\}\nonumber \\
		& = \max_{a^i_t} \{ \qd(\bar{\mat{R}}^i_t+\bar{\mat{Z}}^i_{t+1};\left[\begin{array}{c}
		\hv^i_t \\ a^i_t \\ \vc{f}_t
		\end{array}\right])+(\bar{\vc{b}}^{i\prime}_t+\bar{\vc{z}}^{i\prime}_{t+1})\left[\begin{array}{c}
		\hv^i_t \\ a^i_t \\ f_t
		\end{array}\right]\nonumber \\&+\bar{\vc{c}}^i_t+\bar{o}^i_{t+1}\}.
		\label{quad-a}
		\end{align}
		The above equation is quadratic with respect to $a^i_t$ and therefore, if $(\bar{\mat{R}}^i_t+\bar{\mat{Z}}^i_{t+1})_{a^i,a^i}$ is negative semi definite, the maximum value is achieved when the gradient of the above equation with respect to $a^i_t$ is zero.
		\begin{subequations}
			\begin{align}
			&2(\bar{\mat{R}}^i_t+\bar{\mat{Z}}^i_{t+1})_{a^i,a^i}a^i_t+2(\bar{\mat{R}}^i_t+\bar{\mat{Z}}^i_{t+1})_{a^i,\hv^if}\left[\begin{array}{c}
			\hv^i_t \\ f_t
			\end{array}\right]\nonumber \\&
			\quad +(\bar{\vc{b}}^i_t+\bar{\vc{z}}^i_{t+1})_{a^i}=0
			\\& \Rightarrow a^i_t=-(\bar{\mat{R}}^i_t+\bar{\mat{Z}}^i_{t+1})_{a^i,a^i}^{-1}((\bar{\mat{R}}^i_t+\bar{\mat{Z}}^i_{t+1})_{a^i,\hv^if}\left[\begin{array}{c}
			\hv^i_t \\ f_t
			\end{array}\right]
			\nonumber	\\&\hspace{5cm}   +\frac{1}{2}(\bar{\vc{b}}^i_t+\bar{\vc{z}}^i_{t+1})_{\vc{a}^i}).
			\end{align}
		\end{subequations}
		
		Finally, we can derive the best response strategy of player $i$ to be $\gamma^i_t(\cdot|\hv^i_t)=\delta(\vc{a}^i_t-\mat{L}^i_t\hv^i_t-\vc{m}^i_t)$ where
		\begin{subequations}
			\begin{align}
			& \mat{L}^i_t=-(\bar{\mat{R}}^i_t+\bar{\mat{Z}}^i_{t+1})_{a^i,a^i}^{-1}(\bar{\mat{R}}^i_t+\bar{\mat{Z}}^i_{t+1})_{a^i,\hv^i} \label{Lt}\\
			& \vc{m}^i_t= -(\bar{\mat{R}}^i_t+\bar{\mat{Z}}^i_{t+1})_{a^i,a^i}^{-1}((\bar{\mat{R}}^i_t+\bar{\mat{Z}}^i_{t+1})_{a^i,f}f_t \nonumber\\& \hspace{3.5cm}+\frac{1}{2}(\bar{\vc{b}}^i_t+\bar{\vc{z}}^i_{t+1})_{\vc{a}^i}).
			\end{align}
		Note that we have $\vc{m}^i_t=\mat{M}^i_tf_t+\bar{m}^i_t$, where
	\begin{align}
\mat{M}^i_t&=-(\bar{\mat{R}}^i_t+\bar{\mat{Z}}^i_{t+1})_{a^i,a^i}^{-1}(\bar{\mat{R}}^i_t+\bar{\mat{Z}}^i_{t+1})_{a^i,f}\\
\bar{m}^i_t&=-\frac{1}{2}(\bar{\mat{R}}^i_t+\bar{\mat{Z}}^i_{t+1})_{a^i,a^i}^{-1}(\bar{\vc{b}}^i_t+\bar{\vc{z}}^i_{t+1})_{\vc{a}^i}.
\end{align}
\label{strategy}
		\end{subequations}		
		By substituting the best response action in the reward-to-go equation \eqref{quad-a}, we have the following final step of the proof.
		\begin{align}
		J^i_t&(\hv^i_t,\mat{\Sigma}_t,\mat{E}_t,\vc{f}_t)=\qd(\mat{Z}^i_t;\left[\begin{array}{c}
		\hv^i_t\\\vc{f}_t
		\end{array}\right])+\vc{z}^{i\prime}_t\left[\begin{array}{c}
		\hv^i_t\\\vc{f}_t
		\end{array}\right]+o^i_t,
		\end{align}
		where
		\begin{subequations}
			\begin{align}
			&  \mat{Z}^i_t=\hat{\mat{T}}^{i\prime}_t(\bar{\mat{R}}^i_t+\bar{\mat{Z}}^i_{t+1})\hat{\mat{T}}^i_t \\
			&\hat{\mat{T}}^i_t=\left[\begin{array}{cc}
			\mat{I}_{N_v} & \mat{0}\\
			\mat{L}^i_t & \mat{M}^i_t\\
			\mat{0} & \mat{I}_{N_vN}
			\end{array}\right]\\
			&\vc{z}^{i\prime}_t=2 \hat{\vc{m}}^{i\prime}_t(\bar{\mat{R}}^i_t+\bar{\mat{Z}}^i_{t+1})\hat{\mat{T}}^i_t +(\bar{\vc{b}}^{i\prime}_t+\bar{\vc{z}}^{i\prime}_{t+1})\hat{\mat{T}}^i_t
			\\& \hat{\vc{m}}^i_t=\left[\begin{array}{c}
			\mat{0}\\
			\bar{m}^i_t\\
			\mat{0}
			\end{array}\right]
			\\& o^i_t=\hat{\vc{m}}^{i\prime}_t(\bar{\mat{R}}^i_t+\bar{\mat{Z}}^i_{t+1})\hat{\vc{m}}^i_t+(\bar{\vc{b}}^{i\prime}_t+\bar{\vc{z}}^{i\prime}_{t+1})\hat{\vc{m}}^i_t+\bar{\vc{c}}^i_t+\bar{o}^i_{t+1}.
			\end{align}
			\label{Zt}
		\end{subequations}
		Note that in order to derive the  $\gamma^i_t$ strategy matrix and vector, $\mat{L}^i_t$ and $\vc{m}^i_t$, we need to know $\mat{L}^{-i}_t$ and $\vc{m}^{-i}_t$. Clearly, the same is true for calculating $\mat{L}^{-i}_t$ and $\vc{m}^{-i}_t$. On the other hand, some of  the quantitites used in the proof, like $\hat{\mat{G}}^i_{t+1}$, require $\mat{L}^i_t$ and $\vc{m}^i_t$ to be evaluated.  Therefore, we have a fixed point equation over  $\mat{L}_t$ and $\vc{m}_t$.
		
		Note that we have such linear solution only if the matrix $(\bar{\mat{R}}^i_t+\bar{\mat{Z}}^i_{t+1})_{a^i,a^i}$ is invertible and negative semidefinite for all $i \in \cN$.
	\end{IEEEproof}

We conclude the proof of the theorem by noting that in Lemma \ref{retogoquad}, we proved that the reward to go is a quadratic function of $\left[\begin{array}{c}
	\hv^i_t\\\vc{f}_t
\end{array}\right]$ and as a result and throughout the proof, we  derived equation \eqref{strategy} for the best response strategy of player $i$. Equation \eqref{strategy} indicates that the linear strategies in terms of $\left[\begin{array}{c}
\hv^i_t\\\vc{f}_t
\end{array}\right]$  form equilibria of the game and the theorem is proved.
	
	\bibliographystyle{IEEEtran}
	\optv{submission}{\bibliography{achilleas18abrv,achilleas18_own,achilleas18_control,Nasimeh}}

\begin{thebibliography}{10}
\providecommand{\url}[1]{#1}
\csname url@samestyle\endcsname
\providecommand{\newblock}{\relax}
\providecommand{\bibinfo}[2]{#2}
\providecommand{\BIBentrySTDinterwordspacing}{\spaceskip=0pt\relax}
\providecommand{\BIBentryALTinterwordstretchfactor}{4}
\providecommand{\BIBentryALTinterwordspacing}{\spaceskip=\fontdimen2\font plus
\BIBentryALTinterwordstretchfactor\fontdimen3\font minus
  \fontdimen4\font\relax}
\providecommand{\BIBforeignlanguage}[2]{{%
\expandafter\ifx\csname l@#1\endcsname\relax
\typeout{** WARNING: IEEEtran.bst: No hyphenation pattern has been}%
\typeout{** loaded for the language `#1'. Using the pattern for}%
\typeout{** the default language instead.}%
\else
\language=\csname l@#1\endcsname
\fi
#2}}
\providecommand{\BIBdecl}{\relax}
\BIBdecl

\bibitem{HeAn19}
N.~Heydaribeni and A.~Anastasopoulos, ``Linear equilibria for dynamic {LQG}
  games with asymmetric information and dependent types,'' \emph{IEEE
  Conference on Decision and Control (CDC)}, 2019.

\bibitem{OsRu94}
M.~J. Osborne and A.~Rubinstein, \emph{A course in game theory}.\hskip 1em plus
  0.5em minus 0.4em\relax MIT press, 1994.

\bibitem{FuTi91}
D.~Fudenberg and J.~Tirole, \emph{Game theory}.\hskip 1em plus 0.5em minus
  0.4em\relax MIT press, 1991.

\bibitem{watson2017general}
J.~Watson, ``A general, practicable definition of perfect bayesian
  equilibrium,'' \emph{unpublished draft}, 2017.

\bibitem{maskin2001markov}
E.~Maskin and J.~Tirole, ``Markov perfect equilibrium: I. observable actions,''
  \emph{Journal of Economic Theory}, vol. 100, no.~2, pp. 191--219, 2001.

\bibitem{OuTaTe17}
Y.~Ouyang, H.~Tavafoghi, and D.~Teneketzis, ``Dynamic games with asymmetric
  information: Common information based perfect bayesian equilibria and
  sequential decomposition,'' \emph{IEEE Trans.~Automatic Control}, vol.~62,
  no.~1, pp. 222--237, Jan 2017.

\bibitem{vasal2018systematic}
D.~Vasal, A.~Sinha, and A.~Anastasopoulos, ``A systematic process for
  evaluating structured perfect bayesian equilibria in dynamic games with
  asymmetric information,'' \emph{IEEE Transactions on Automatic Control},
  vol.~64, no.~1, pp. 81--96, 2018.

\bibitem{mahajan2015sufficient}
A.~Mahajan and A.~Nayyar, ``Sufficient statistics for linear control strategies
  in decentralized systems with partial history sharing,'' \emph{IEEE
  Transactions on Automatic Control}, vol.~60, no.~8, pp. 2046--2056, 2015.

\bibitem{yuksel2009stochastic}
S.~Yuksel, ``Stochastic nestedness and the belief sharing information
  pattern,'' \emph{IEEE Transactions on Automatic Control}, vol.~54, no.~12,
  pp. 2773--2786, 2009.

\bibitem{tavafoghi2018unified}
H.~Tavafoghi, Y.~Ouyang, and D.~Teneketzis, ``A unified approach to dynamic
  decision problems with asymmetric information-part ii: Strategic agents,''
  \emph{arXiv preprint arXiv:1812.01132}, 2018.

\bibitem{VaAn16a}
D.~Vasal and A.~Anastasopoulos, ``Signaling equilibria for dynamic {LQG} games
  with asymmetric information,'' in \emph{Proc.~IEEE Conf. on Decision and
  Control}, Dec. 2016, pp. 6901--6908.

\bibitem{abreu1990toward}
D.~Abreu, D.~Pearce, and E.~Stacchetti, ``Toward a theory of discounted
  repeated games with imperfect monitoring,'' \emph{Econometrica: Journal of
  the Econometric Society}, pp. 1041--1063, 1990.

\bibitem{BiHiWe92}
S.~Bikhchandani, D.~Hirshleifer, and I.~Welch, ``A theory of fads, fashion,
  custom, and cultural change as informational cascades,'' \emph{Journal of
  political Economy}, pp. 992--1026, 1992.

\bibitem{ho1972team}
Y.-C. Ho \emph{et~al.}, ``Team decision theory and information structures in
  optimal control problems--part i,'' \emph{IEEE Transactions on Automatic
  control}, vol.~17, no.~1, pp. 15--22, 1972.

\bibitem{KuVa86}
P.~R. Kumar and P.~Varaiya, \emph{Stochastic systems: estimation,
  identification, and adaptive control}.\hskip 1em plus 0.5em minus 0.4em\relax
  Englewood Cliffs, NJ: Prentice-Hall, 1986.

\bibitem{witsenhausen1968counterexample}
H.~S. Witsenhausen, ``A counterexample in stochastic optimum control,''
  \emph{SIAM Journal on Control}, vol.~6, no.~1, pp. 131--147, 1968.

\bibitem{basar1978two}
T.~Ba\c{s}ar, ``Two-criteria {LQG} decision problems with one-step delay
  observation sharing pattern,'' \emph{Information and Control}, vol.~38,
  no.~1, pp. 21--50, 1978.

\bibitem{altman2009stochastic}
E.~Altman, V.~Kambley, and A.~Silva, ``Stochastic games with one step delay
  sharing information pattern with application to power control,'' in
  \emph{2009 International Conference on Game Theory for Networks}.\hskip 1em
  plus 0.5em minus 0.4em\relax IEEE, 2009, pp. 124--129.

\bibitem{BiAn18}
I.~Bistritz and A.~Anastasopoulos, ``Characterizing non-myopic information
  cascades in {B}ayesian learning,'' in \emph{Proc.~IEEE Conf. on Decision and
  Control}, Miami Beach, FL, July 2018.

\bibitem{BiHeAn19j}
I.~Bistritz, N.~Heydaribeni, and A.~Anastasopoulos, ``Do {I}nformational
  {C}ascades {H}appen with {N}on-myopic {A}gents?'' \emph{arXiv preprint
  arXiv:1905.01327}, 2019.

\bibitem{HeBiAn19}
N.~Heydaribeni, I.~Bistritz, and A.~Anastasopoulos, ``{I}nformational cascades
  can be avoided with non-myopic agents,'' \emph{57th Annual Allerton
  Conference}, 2019.

\bibitem{GuNaLaBa14b}
A.~Gupta, A.~Nayyar, C.~Langbort, and T.~Ba{\c s}ar, ``Common information based
  {M}arkov perfect equilibria for linear-{G}aussian games with asymmetric
  information,'' \emph{SIAM Journal on Control and Optimization}, vol.~52,
  no.~5, pp. 3228--3260, 2014.

\bibitem{crawford1982strategic}
V.~P. Crawford and J.~Sobel, ``Strategic information transmission,''
  \emph{Econometrica: Journal of the Econometric Society}, pp. 1431--1451,
  1982.

\bibitem{farokhi2014gaussian}
F.~Farokhi, A.~M. Teixeira, and C.~Langbort, ``Gaussian cheap talk game with
  quadratic cost functions: When herding between strategic senders is a
  virtue,'' in \emph{2014 American Control Conference}.\hskip 1em plus 0.5em
  minus 0.4em\relax IEEE, 2014, pp. 2267--2272.

\bibitem{kamenica2011bayesian}
E.~Kamenica and M.~Gentzkow, ``Bayesian persuasion,'' \emph{American Economic
  Review}, vol. 101, no.~6, pp. 2590--2615, 2011.

\bibitem{sayin2018dynamic}
M.~O. Sayin and T.~Ba{\c{s}}ar, ``Dynamic information disclosure for
  deception,'' in \emph{2018 IEEE Conference on Decision and Control
  (CDC)}.\hskip 1em plus 0.5em minus 0.4em\relax IEEE, 2018, pp. 1110--1117.

\end{thebibliography}
	\optv{arxiv}{

}
	\vspace*{-0.5cm}
	\begin{IEEEbiography}[{\includegraphics[width=1in,height=1.25in,clip,keepaspectratio]{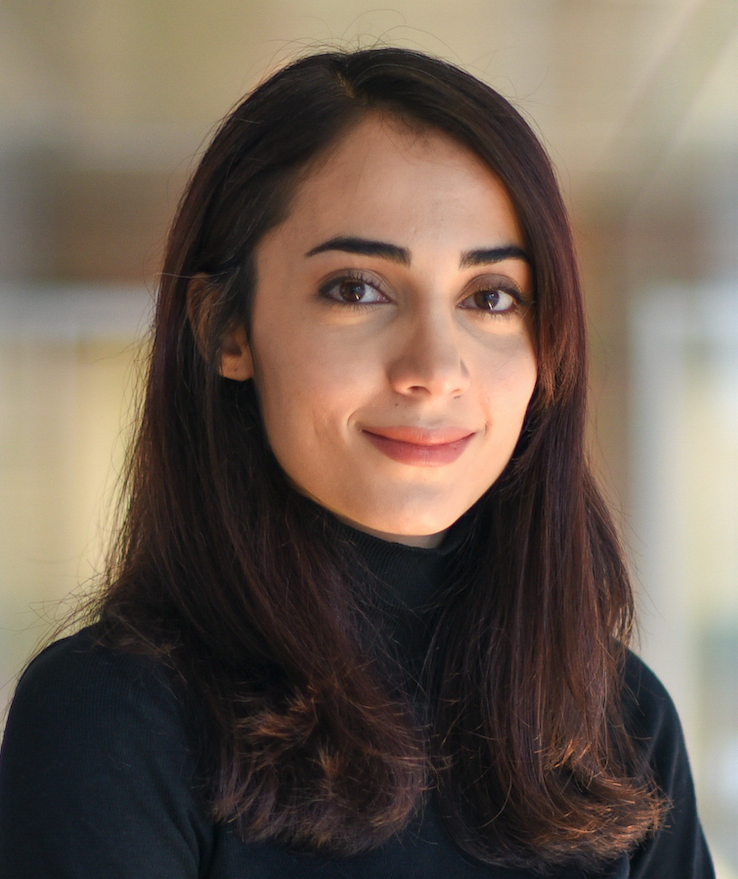}}]{Nasimeh Heydaribeni}
	received her B.S. and M.S. degrees in Electrical Engineering at Sharif University of Technology, Tehran, Iran, in 2015 and 2017, respectively. She is currently pursuing her Ph.D. degree in Electrical Engineering and Computer Science at University of Michigan, Ann Arbor. Her research interests are game theory and its applications in networked systems with emphasis on mechanism design, dynamic games with asymmetric information and information design.
\end{IEEEbiography}
\vfill

	\vspace*{-0.5cm}

\begin{IEEEbiography}[{\includegraphics[width=1in,height=1.25in,clip,keepaspectratio]{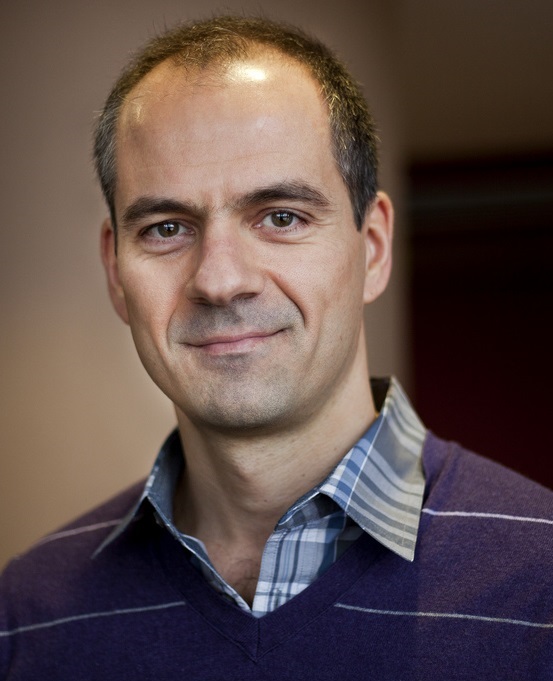}}]{Achilleas Anastasopoulos}
(S'97-M'99-SM'13) was born in Athens, Greece in 1971. He received the Diploma
in Electrical Engineering from the National Technical University of
Athens, Greece in 1993, and the M.S.\ and Ph.D.\ degrees in Electrical
Engineering from University of Southern California in 1994 and 1999,
respectively. He is currently an Associate Professor at the University of
Michigan, Ann Arbor, Department of Electrical Engineering and Computer
Science.

His research interests lie in the general area of communication and information theory,
with emphasis in channel coding and multi-user channels;
control theory with emphasis in decentralized stochastic control and its
connections to communications and information theoretic problems;
analysis of dynamic games and mechanism design for resource allocation on networked systems.

He is the co-author
of the book \emph{Iterative Detection: Adaptivity, Complexity Reduction,
and Applications,} (Reading, MA: Kluwer Academic, 2001).

Dr.\ Anastasopoulos is the recipient of the ``Myronis Fellowship'' in
1996 from the Graduate School at the University of Southern California,
the NSF CAREER Award in 2004,
and was a co-author for the paper that received the best student paper award in ISIT 2009.
He served as a technical program committee member for 
ICC 2003, 2015--2018; 
Globecom 2004, 2012; 
VTC 2007, 2014, 2015; 
ISIT 2015,
SPAWC 2018,
and is currently serving as the TPC co-Chair for the Communication Theory Symposium, ICC'21.
He was an associate editor for the IEEE Transactions on Communications in 2003--2008.
%

%
\end{IEEEbiography}
\vfill

\end{document}